\documentclass[10pt,journal,compsoc]{IEEEtran}
\usepackage{amsmath,amsfonts}
\usepackage{algorithmic}
\usepackage{algorithm}
\usepackage{array}
\usepackage[caption=false,font=normalsize,labelfont=sf,textfont=sf]{subfig}
\usepackage{textcomp}
\usepackage{stfloats}
\usepackage{url}
\usepackage{verbatim}
\usepackage{graphicx}
\usepackage{cite}
\usepackage{etoolbox}
\usepackage{xcolor}

\newbool{showcorrections}
\booltrue{showcorrections}
\boolfalse{showcorrections}

\newcommand{\edit}[2]{%
  \ifbool{showcorrections}%
    {\color{red}#1\color{blue}#2\color{black}}%
    {#2}%
}

\newbool{showcorrections_minor_2024_02}
\booltrue{showcorrections_minor_2024_02}
\boolfalse{showcorrections_minor_2024_02}

\newcommand{\editMinor}[2]{%
  \ifbool{showcorrections_minor_2024_02}%
    {\color{black}\color{red}#1\color{blue}#2\color{black}}%
    {\color{black}#2\color{black}}%
}

\newcommand{\mysize}{0.2}

\begin{document}

\title{De-cluttering Scatterplots with Integral Images}

\author{Hennes~Rave, Vladimir~Molchanov, and Lars~Linsen
\thanks{All authors are with the University of Münster, Germany. \protect\\
E-mail: \{hennes.rave\textbar molchano\textbar linsen\}@uni-muenster.de.}
\thanks{Manuscript received xxxxx; revised xxxxx.}}

\markboth{IEEE Transactions on Visualization and Computer Graphics}%
{Rave \MakeLowercase{\textit{et al.}}: De-cluttering Scatterplots with Integral Images}


\maketitle

\begin{abstract}
Scatterplots provide a visual representation of bivariate data (or 2D embeddings of multivariate data) that allows for effective analyses of data dependencies, clusters, trends, and outliers. Unfortunately, classical scatterplots suffer from scalability issues, since growing data sizes eventually lead to overplotting and visual clutter on a screen with a fixed resolution, which hinders the data analysis process. We propose an algorithm that compensates for irregular sample distributions by a smooth transformation of the scatterplot's visual domain. Our algorithm evaluates the scatterplot's density distribution to compute a regularization mapping based on integral images of the rasterized density function. The mapping preserves the samples' neighborhood relations. Few regularization iterations suffice to achieve a nearly uniform sample distribution that efficiently uses the available screen space. We further propose approaches to visually convey the transformation that was applied to the scatterplot and compare them in a user study. We present a novel parallel algorithm for fast GPU-based integral-image computation, which allows for integrating our de-cluttering approach into interactive visual data analysis systems.
\end{abstract}


\section{Introduction}

Visual representation of multidimensional data has been and remains a challenging task, which becomes more demanding as the dimensionality and size of datasets steadily grow. \emph{Scatterplots} are an effective and thus widely used method for visualizing multidimensional data in a 2D domain by relating pairs of data dimensions. Scatterplots reveal the structure of the data including outliers, clusters, patterns, and tendencies.

When rendering scatterplots within a visual domain on a screen with a fixed resolution, growing numbers of data samples eventually lead to occlusion and \textit{overplotting}\editMinor{. Thus, scatterplots suffer from a data size scalability issue, as the growing numbers of plotted data points}{, } negatively affect\editMinor{}{ing } user perception and hinder\editMinor{}{ing } visual data analysis. In particular, it becomes difficult to visually estimate the number of samples and their density in cluttered regions\editMinor{, which prevents effective cluster analysis in occluded layouts}{}. Moreover, access to individual data samples is restricted, which impedes user exploration, e.g., of image data collections~\cite{Eler09}.

Several approaches were proposed to alleviate the overplotting issue. A common strategy is to adjust the transparency of the displayed samples to improve the visibility of the local density of the points' distribution~\cite{Micallef17}. Alternatively, the number of rendered elements can be reduced by applying a down-sampling strategy to the dataset~\cite{Dix02, Hu20, Yuan2021}. Such approaches change the appearance of the presentation. Another broad class of methods uses \textit{spatial distortions} of the representation to reduce overplotting. These techniques may remap samples to pixels or distort the visualization domain for more efficient use of the screen space. \editMinor{Many of the existing approaches require non-trivial computations or involved data analysis to be effective for general multidimensional datasets.}{}

We propose a numerical method for a smooth iterative deformation of the scatterplot domain aiming at optimizing the available plotting space usage. In each iteration, the per-pixel deformation is constructed based on a set of density integrals, so-called \textit{integral images (InIms)}\editMinor{, which are efficiently computed on the GPU}{}. Here, a crucial difference of the proposed method from existing deformation approaches is that the integrals are not restricted to any local neighborhood but rather characterize global density distribution. \edit{}{Therefore, the displacement of each sample depends on the global data distribution encoded in InIms rather than the samples' distribution in its neighborhood. } As \edit{the }{a } result, no expensive collision detection is needed and large datasets \edit{}{(i.e., datasets with the number of points close to the screen resolution or even beyond) }
can be regularized at highly interactive rates. The transformed scatterplot has a nearly regular sample distribution, significantly mitigating the overplotting issues, and making large amounts of samples visible and manageable. 

Our proposed deformation preserves essential properties of the original scatterplot such as \textit{neighborhood relations} of the displayed samples, including their \edit{}{local } ordering, which does not automatically hold in parameter-sensitive smoothing-based regularizations. In contrast to sampling approaches, the proposed algorithm is \textit{deterministic} and \textit{preserves all data samples} in the visualization domain. We show that, in comparison to opacity-based approaches, the alternative regularized view allows the user to better perceive the density and quantity of data samples, more easily analyze class-cluster relations, and more easily retrieve further information about individual samples. 

Our regularized scatterplots are meant to complement the original scatterplots. While the original scatterplots convey the global data distribution including clusters and outliers, our deformed versions give access to detailed structures that were occluded in the original scatterplot.
We support continuous transitions between the two views, 
following the \textit{reconfigure} interaction principle~\cite{yi07}.
\editMinor{\edit{In addition }{Additionly}}{Additionally}, we propose several approaches for \textit{conveying initial distribution} of samples of the original scatterplot in the deformed scatterplots. We present grid lines, background texture, and contour lines and compare them with each other.

The primary target applications of our approach are within interactive visual analyses, where the user controls the desired level of deformation. \editMinor{Our proposed visual space deformation can be embedded into any visual analysis system, where it is complemented with other interaction mechanisms such as zooming, panning, brushing, or further selection methods. }{} Other interactions benefit from our regularization of the samples' distribution. For example, less zooming would be required after regularization, and individual samples would be more accessible, e.g., for selections via brushing, clicking, or hovering.
The \textit{interactivity} constraints of the visualization system demand high computational efficiency of our proposed algorithm and good computational scalability. We satisfy these requirements by proposing a novel algorithm that enables fast GPU implementation of all necessary computations.

Our main contributions can be summarized as follows:
\begin{itemize}
    \item We provide a novel deterministic technique for de-cluttering scatterplots using {integral images} (InIms), which substantially improves stability, convergence, and efficiency of a prior algorithm~\cite{Molchanov20_wscg}. \editMinor{}{We compare our approach to the state of the art for visual clutter reduction in scatterplots. Our GPU implementation is made publicly available.}
    \editMinor{\item We present a fast GPU implementation of the algorithm that allows for embedding it into any interactive visual data analysis system and for combining it with common interactive data exploration tools. Our GPU implementation is made publicly available. }{}
    \item We apply the algorithm for a smooth transformation of the visual scatterplot domain\editMinor{resulting in a nearly-regular sample distribution with preserved samples' neighborhoods}{}, where the level of deformation can be \editMinor{easily}{} adjusted \editMinor{interactively}{} by the user.
    \item We propose different visual encodings to convey the amount of local deformations, information about the cluster structure, and the samples' density in the deformed plots\editMinor{, which support a reliable analysis of the original data after transformation.}{. We evaluate their effectiveness in a controlled user study.}
    \editMinor{
    \item We evaluate in a controlled user study how effective the visual encodings of local deformations are in conveying the structures of the undeformed scatterplot, i.e., how effective they are in supporting common analysis tasks on the deformed scatterplot.
    \item We compare regularized layouts computed by our proposed method to the state of the art, i.e., to the most similar, popular, or recent techniques for visual clutter reduction in scatterplots.}{}
\end{itemize}

\edit{Concerning the structure of the paper, we first provide an overview and discuss related approaches for visual clutter reduction in Section~\ref{sec:related_work}. }{ } 
\editMinor{Our deformation approach is based on the concept of InIms.}{} We provide respective background information and introduce required terms and notations in Section~\ref{sec:background}. 
In Section~\ref{sec:method}, we present the basic deformation formula as presented by Molchanov and Linsen~\cite{Molchanov20_wscg}, discuss its limitations\edit{}{,} and provide the key steps of our deformation algorithm including GPU implementation details. 
Section~\ref{sec:preservation} is dedicated to the proposed visual encodings, which serve to compensate for the local samples' density information typically lost after deformation.
Results of our numerical tests and details on the conducted user study are provided in Section~\ref{sec:results}. \edit{Finally, we summarize our work in Section~\ref{sec:conclusion}.}{}

\section{Related Work}
\label{sec:related_work}

\textit{Scatterplots} are arguably the most commonly used visualization method for data sets with more than one numerical dimension and have a long-standing history~\cite{Friendly05}. 
Traditional scatterplots relate two data dimensions by drawing data samples as points in a 2D Cartesian coordinate system. Frequently, the original data are multidimensional and some dimensionality reduction technique precedes the visualization step. The two scatterplot dimensions can, thus, be selected from the given set of data attributes or may represent a linear combination of them like in principal component analysis~\cite{Jolliffe86} or, more generally, in star-coordinates projections~\cite{Kandogan00, Kandogan01}. Moreover, they could also be a 2D embedding of the multidimensional data using non-linear dimensionality reduction methods such as multidimensional scaling (MDS)~\cite{Kruskal64, Kruskal78}, t-distributed stochastic neighbor embedding (t-SNE)~\cite{Maaten08} or uniform manifold approximation and projection (UMAP)~\cite{mcinnes2018umap}. An extensive user study on the combination of various dimensionality reduction techniques with scatterplot representations for data exploration tasks was performed by Sedlmair et al.~\cite{Sedlmair13}.

While conventional scatterplots render data samples as points, different approaches propose to incorporate additional information using, e.g., \textit{glyphs} or summary visualizations. Chan et al.~\cite{Chan10} enhanced scatterplots with sensitivity coefficients representing local correlations of data. Janetzko et al.~\cite{Janetzko13} used ellipsoid pixel placement for the same purposes. Staib et al.~\cite{Staib16} applied blurring to encode depth of field information.
Such design choices for visual encoding can significantly enhance or degrade their visual quality. Micallef et al.~\cite{Micallef17} proposed a cost function aiming at an automatic optimization of marker size and opacity, aspect ratio, color, and rendering order in scatterplots depending on the analysis task.

Complex and large data cannot be effectively represented in traditional scatterplots due to \textit{overplotting}. For overpopulated plots, there is a need to accentuate the most important data structures, reject or combine less relevant samples, or reduce the local sample density by other means. Recently, Sarikaya and Gleicher~\cite{Sarikaya18} surveyed existing scatterplot designs reasoned by the analysis tasks and data characteristics. A taxonomy of existing methods was developed by Ellis and Dix~\cite{Ellis07}. The authors identified three main strategies for tackling the overplotting issue, which can be characterized as spatial transformation, changing the appearance, and using animations.

\textit{Animations} may help to encode temporally varying data or data with uncertainty~\cite{Feng10}.
Chen et al.~\cite{Chen18} used flickering points for revealing multi-class structures in overplotted scatterplots. Technically, animations can handle relatively large datasets. However, one should consider the time required to show and perceive all the data as well as the cognitive burden~\cite{Ellis07}.

\textit{Appearance change} for clutter reduction may use different sizes for depicting objects~\cite{Li10}. More commonly applied though is the concept of opacity adjustment, which can improve the user's perception of local sample density in overplotted scatterplots. Matejka et al.~\cite{Matejka15} proposed a model for user-driven opacity scaling. Proper sampling of data can also improve the readability of visualization when the data size is large. Bertini et al.~\cite{Bertini06} performed a non-uniform sampling in combination with sample displacement to support user perception of scatterplots.
\emph{Splatterplots} proposed by Mayorga and Gleicher~\cite{Mayorga13} alleviate the overdraw issue in traditional scatterplots by abstracting local sample density and rendering density contours.
Hao et al.~\cite{Hao10} split the scatterplot domain into bins along each spatial dimension and distribute data points within each bin, making individual samples accessible.
Paulovich and Minghim~\cite{Paulovich08} applied a hierarchical approach for visualizing document collection datasets. The proposed technique \emph{HiPP} depicts data at a certain level of detail preserving the similarity of samples and their clusters as inter-object distances.
\textcolor{black}{Among these approaches that change the appearance of the scatterplots, opacity adjustment seems to be the most commonly applied method due to being simple, effective, and not introducing new visual objects. In our user study, we thus compare our method against opacity adjustment, see Section~\ref{sec:user_study}.}

\textit{Spatial transformations} for reducing clutter in scatterplots introduce a distortion of the sample placement. In the distorted view, high-density regions should be expanded, while low-density regions should be contracted to use the available screen space effectively. A taxonomy of distortion-oriented techniques was presented by Leung and Apperley~\cite{Leung94}. Sarkar et al.~\cite{Sarkar93} used a rubber-sheet metaphor for a user-controlled local deformation of the visual domain. Keim et al.~\cite{Keim10} proposed \emph{generalized scatterplots} trading off overlap and distortion errors. The method combines a linear domain distortion technique~\cite{Keim03} with a pixel displacement interactively controlled by the user. Recently, Raidou et al.~\cite{Raidou19} computed a space-filling transformation that maps data samples to free pixels. Vollmer and D\"{o}llner~\cite{Vollmer20} proposed a collision detection algorithm and 2.5D layout for alleviating occlusion. Cutura et al.~\cite{Cutura21, Cutura22} resolved collisions on space-filling curves for gridifying the scatterplot layout. Hilasaca et al.~\cite{Hilasaca23} remove overlapping of glyphs by creating dummy points. Liu et al.~\cite{Liu18} applied constrained MDS for placing data items on a grid.

Local reduction of the samples' occlusion can be achieved by using \emph{virtual lenses} or other similar \emph{focus + context} techniques. \textit{Fisheye views} proposed by Furnas~\cite{Furnas86} provide a smooth integration of two levels of details. \textit{JellyLenses} developed by Pindat et al.~\cite{Pindat12} dynamically adapt their geometry to the content in the focus. The application of virtual lenses to scatterplots and parallel coordinates plots was discussed by Ellis et al.~\cite{Ellis05, Ellis06}. Tominski et al.~\cite{Tominski17} presented a taxonomy of virtual interactive lenses according to the types of data and user tasks.
\textcolor{black}{Local reduction methods come with the drawback of requiring rather intense user interactions when exploring the entire visual space and imposing some cognitive load on the user when trying to compare different regions to each other.}

In our work, we propose a novel global domain deformation technique, which aims at reducing the scatterplot occlusion. 
The deformation map computation is based on \emph{summed-area tables} or \emph{integral images} (InIms) originally introduced by Crow~\cite{Crow84}. Viola and Jones~\cite{Viola02} applied InIms to object detection in image analysis. Ehsan et al.~\cite{Ehsan15} addressed the problem of efficient parallel computation of InIms. Singhal et al.~\cite{Singhal12} discussed the calculation of InIms on the embedded GPU using an OpenGL shader model. The proposed implementation uses a multi-pass 2D reduction technique performed in a fragment shader. Nowadays, many libraries like OpenCV~\cite{opencv} provide out-of-the-box functionality for efficient InIm computation. Reinbold and Westermann~\cite{Reinbold21} presented a hierarchical approach for computing summed volume tables for \edit{sparce }{sparse } volumes. An extension to the classical InIms, which accumulate pixel values in the left-top corner of the input texture, are InIms tilted by $45^\circ$. Lienhart et al.~\cite{Lienhart02, Lienhart03} presented a computation of tilted InIms on the CPU in two passes over all pixels. The authors used rotated InIms for the calculation of additional Haar-like features serving for a more robust detection of objects. Barczak et al.~\cite{Barczak06} extended the approach for $26.5^\circ$ and $63.5^\circ$ angles of rotation as well as constructed approximations for arbitrary angles. Computation of InIms at arbitrary angles was studied by Chin et al.~\cite{Chin08}.

Molchanov and Linsen~\cite{Molchanov20_wscg} used InIms for computing virtual lenses and pseudo-cartograms. Their deformation formula cannot be used to achieve a nearly uniform distribution of samples in general cases, since the ideal regular distribution of samples is not a fixed point of the transformation. We correct the mapping presented in~\cite{Molchanov20_wscg}, so that an iterative regularization of arbitrary layout converges to the desired state, is stable, avoids overlapping of mapped subareas, and has better performance scalability in terms of data size.

Spatial distortion changes the original distances between visualized elements, which may lead to misinterpretations of the data's structure by the user. Therefore, it is important to \textit{inform the user about applied distortions} and so \textit{relate the regularized layout back to its original representation}. Carpendale et al.~\cite{Carpendale95} presented information about the properties of local map transformations via superimposed regular grids. In our work, we introduce different techniques for depicting spatial distortion including grids and color encoding of the background, which we evaluate within a user study in Section~\ref{sec:user_study}.

\section{Background}\label{sec:background}

In this section, we provide basic concepts and necessary technical details of the existing algorithms, which we improve, adapt, and experiment with in our work. We start with a given real-valued density texture $d(i,\,j)$, whose construction is application-specific. The density depicts local weights, which may be interpreted as the amount of information or level of importance of the current locus. We propose a proper calculation of $d(i,\,j)$ for de-cluttering scatterplots later in Section~\ref{sec:densityfield}. For easier reference to prior work on InIms (e.g., \cite{Molchanov20_wscg}), we use the same notations, where applicable.

\subsection{Integral Images}\label{sec:inims}

For a given real-valued texture $d(i,\,j)$, an \textit{InIm} is another real-valued texture $\alpha(i,\,j)$ of the same resolution, which is computed by summing up all values of the input texture $d(i,\,j)$ over its top-left corner up to the position of pixel $(i,\,j)$. We will also need a set of further sums of pixel values computed over the bottom-left, bottom-right, and top-right corners of the visualization domain, i.e.,
\begin{align*}\label{alpha}
\alpha(i,\,j) &= \sum\limits_{i'\leq i} \sum\limits_{j'\leq j} d(i',\,j'), &
\beta(i,\,j) &= \sum\limits_{i'\leq i} \sum\limits_{j'>j} d(i',\,j') \\
\gamma(i,\,j) &= \sum\limits_{i'> i} \sum\limits_{j'> j} d(i',\,j'), &
\delta(i,\,j) &= \sum\limits_{i'> i} \sum\limits_{j'\leq j} d(i',\,j').
\end{align*}
Note that the areas of summations form a partition of the texture domain, i.e., they do not intersect and their union is the complete texture domain.
Thus, $\alpha+\beta+\gamma+\delta\equiv C$ holds for all pixels, where 
$C =\sum d(i,\,j)$ is the sum of all pixel values of the given texture. 

Next, we define \textit{tilted InIms} with a tilt of $45^\circ$ by
\begin{align*}
\alpha_t(i,\,j) &= \sum\limits_{\substack{i'+j'\leq i+j\\i'-j'\geq i-j}} d(i',\,j'), &
\beta_t(i,\,j)  &= \sum\limits_{\substack{i'+j'\leq i+j\\i'-j'<i-j}} d(i',\,j'),\\
\gamma_t(i,\,j) &= \sum\limits_{\substack{i'+j'> i+j\\i'-j'<i-j}} d(i',\,j'), &
\delta_t(i,\,j) &= \sum\limits_{\substack{i'+j'> i+j\\i'-j'\geq i-j}} d(i',\,j').
\end{align*}
Analogously to InIms, the summation areas form a partition of the domain, and the equation $\alpha_t+\beta_t+\gamma_t+\delta_t\equiv C$ holds.
InIms $\alpha$ and $\alpha_t$ can be computed using standard functions provided by many image processing libraries. The other tables, i.e., $\beta$, $\gamma$, $\delta$, and their tilted versions, can be computed as standard InIms of the density texture when rotated by $90^\circ$, $180^\circ$, and $270^\circ$, correspondingly.

Thus, taking into account these two relations between the introduced integral tables, there are six linearly independent textures, namely $\alpha$, $\beta$, $\gamma$, $\alpha_t$, $\beta_t$, and $\gamma_t$. InIms provide a pixel-centered description of the density distribution on the global scope. Based on their values, it is possible to globally define a density-equalizing mapping, which automatically preserves the ordering of samples.

\subsection{Domain Transformation}
\label{sec:meshmapping}

Texture $d(i,\,j)$ represents density values. High values indicate that more visual space is required locally for optimal representation, i.e., the visualization domain should expand locally. Low-density values identify regions that may be contracted to free more space. However, the density texture itself does not show preferable directions of deformation due to the lack of global scope.

InIms provide integrals of the density distribution over respective domains. Thus, their values represent globally aggregated characteristics of density distribution. Such information can be used for determining the direction and magnitude of local transformation leading to the globally improved (i.e., more uniform) distribution.

Molchanov and Linsen~\cite{Molchanov20_wscg} proposed a global mapping of the following form:
\begin{equation}
\begin{aligned}\label{map}
    t(x,\,y;\,d)= \big(
    &\alpha     \cdot   q_1(x,\,y) & &+ \beta       \cdot   q_2(x,\,y)  & &+ \\
    &\gamma     \cdot   q_3(x,\,y) & &+ \delta      \cdot   q_4(x,\,y)  & &+ \\
    &\alpha_t   \cdot   (x,\,1)    & &+ \beta_t     \cdot   (1,\,y)     & &+ \\
    &\gamma_t   \cdot   (x,\,0)    & &+ \delta_t    \cdot   (0,\,y)\ \big) \big/ (2C).
\end{aligned}
\end{equation}
Here, for each pixel with index $(i,\,j)$ corresponding to the texture coordinates \mbox{$(x,\,y)=2^{-k}\ (i,\,j)$} for a texture of resolution $2^{k}\times 2^{k}$, four anchor points $q_l(x,\,y)$, \mbox{$l=1,\ldots,4$} are defined as follows:
\begin{alignat*}{2}
& y<x:          \quad && q_1(x,\,y)=(1,\,1+y-x),\\
&               \quad && q_3(x,\,y)=(x-y,\,0);\\
& y\geq x:      \quad && q_1(x,\,y)=(1-y+x,\,1),\\
&               \quad && q_3(x,\,y)=(0,\,y-x);\\
& x+y<1:        \quad && q_2(x,\,y)=(x+y,\,0),\\
&               \quad && q_4(x,\,y)=(0,\,x+y);\\
& x+y\geq 1:    \quad && q_2(x,\,y)=(1,\,x+y-1),\\
&               \quad && q_4(x,\,y)=(x+y-1,\,1).
\end{alignat*}

\begin{figure}[!htb]
\centering
\includegraphics[width=\linewidth]{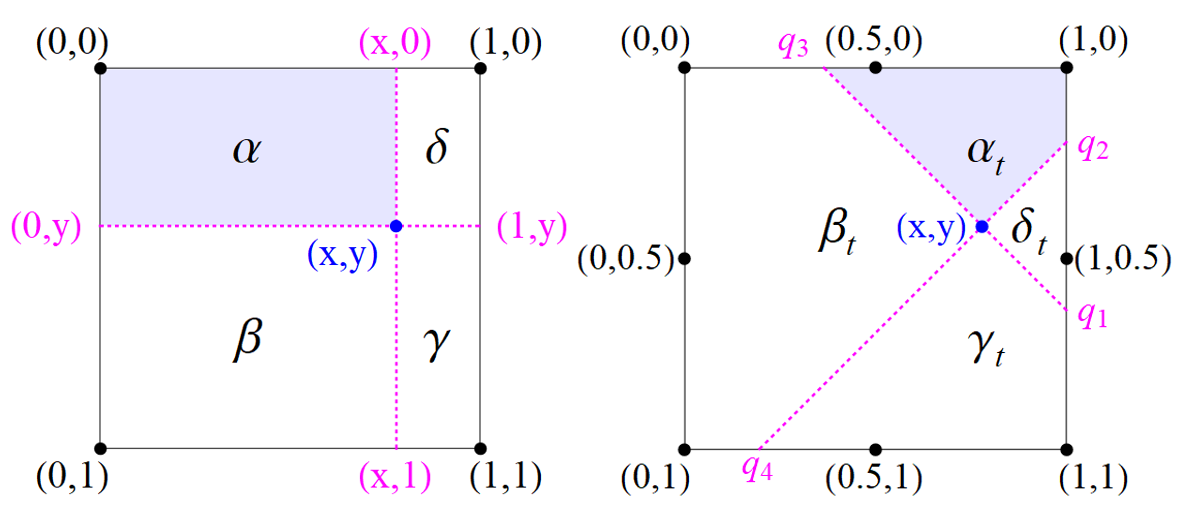}
\caption{\edit{}{Reproduction from Molchanov and Linsen~\cite{Molchanov20_wscg}. Left:~The four InIm coefficients computed at location $(x,y)$ stand for integrals of a density function over respective rectangular regions. Right:~The four additional coefficients can be computed for the same location as integrals over tilted regions.}}\label{fig::ii_both}
\end{figure}

\edit{}{For completeness and self-sufficiency of the exposition, we reproduce Fig.~1 from Molchanov and Linsen~\cite{Molchanov20_wscg}, where the anchor points' positions and important notations are depicted, see Figure~\ref{fig::ii_both}.}

Molchanov and Linsen used mapping~\ref{map} for computing a user-steered virtual lens. The user selects a (multi-component) area of interest in the visual domain. Then, a density function taking constant values inside and outside of the area of interest is constructed. The user adjusts the density value inside the selected area to achieve a desired level of magnification of the region after applying mapping~\ref{map}. Optimal values for the density strongly depend on the shape and position of the area of interest and cannot be computed a priori. Moreover, mapping~\ref{map} is not an identity transformation for a constant texture $d(i,\,j)$ and therefore multiple applications of the mapping to the given distribution may lead to unpredictable results. Therefore, equalizing the distribution using mapping~\ref{map} is not possible.

\section{De-cluttering Scatterplots}\label{sec:method}

Given a scatterplot, our goal is to achieve a uniform distribution of samples in a fully automatic manner. We aim for stable and fast calculations that result in nearly uniform distributions for arbitrary initial configurations. Our first step is to construct a proper density function representing the given scatterplot (Section~\ref{sec:densityfield}). Then, we modify mapping~\ref{map} to act as a converging density-equalizing transformation and solve the stability issue due to its potential singular behavior in the empty regions of scatterplots (Section~\ref{sec:deformation}). We further propose a computationally efficient implementation of the algorithm summarized in Section~\ref{sec:algorithm} on the GPU to allow its application within interactive data analysis and exploration systems (Section~\ref{sec:implementation}).

\subsection{Density Field}
\label{sec:densityfield}

Given a 2D scatterplot with $n$ data samples $\mathbf{z}_i=(x_i,\,y_i)$, $i=1,\ldots,n$.
Without loss of generality, we assume the set of samples belonging to the unit square, i.e., $\{\mathbf{z}_i\}_i\subset [0,\,1]^2$. When rendering the scatterplot, the unit square is mapped to a discrete texture and shown on the screen. We assume that the texture has resolution $2^k\times 2^k$, $k\in\mathbb{N}$. For convenience, we place the origin in the top-left corner of the domain. 

Given the distribution of samples $\{\mathbf{z}_i\}_i$ in the scatterplot, our goal is to define a deformation of the texture space $[0,\,1]^2$, which would result in a more uniform distribution of the samples. The deformation has to be smooth to preserve neighborhood relations between samples, i.e., to avoid mixing and changing the local samples' order.

The first step of the proposed algorithm is the computation of a scalar-valued function, which describes the distribution of data samples in the scatterplot domain. We require this density function to be smooth since it serves as a basis for constructing the deformation map. Such a smooth density distribution can be computed by
\begin{equation*}\label{dr}
d_r(x,\,y) = \sum\limits_{p=1}^n \phi_r(x-x_p, y-y_p),
\end{equation*}
where $\phi_r$ is a smooth radial basis function, e.g., a 2D Gaussian kernel, and $r$ is its dilation parameter. Smaller values of $r$ correspond to more localized contributions from individual samples, while larger values of $r$ result in a smoother density distribution.

In practice, one computes a discrete version in the form of the rasterized density $d_r(i,\,j)$ at pixels $(i,\,j)$. This can be performed using various techniques. One approach is an accumulation of splats in a single-channel float-valued texture as proposed for the construction of continuous scatterplots by Bachthaler and Weiskopf~\cite{Bachthaler08}. Then, larger values of $r$ result in more blurred distributions and longer computational times due to the need to update a higher number of pixel values. Molchanov et al.~\cite{Molchanov13_WSCG} proposed using a spectral algorithm for a fast accumulation of large splats. In our application scenario, we use perfectly isotropic kernels with equal radii, since we assume no variability among the samples. Therefore, we can use a more efficient procedure. The most efficient method for computing density \edit{$d_r(x,\,y)$ }{$d_r(i,\,j)$ } is to restrict the contribution of each data sample to a single pixel of the texture. Then, we convolve the resulting texture with a discrete smoothing kernel such as the 2D Gaussian kernel, and accumulate the smoothed textures.

Depending on the original data and the value of parameter $r$, density texture $d_r(i,\,j)$ may contain vanishing or very small pixel values. When regularizing scattered data, empty regions should be contracted and ideally should disappear. The resulting mapping becomes singular in such regions, which can lead to numerical instabilities and some overlap of mapped subregions. Therefore, to ensure the stability of calculations, we add a global constant value to $d_r(i,\,j)$ for all pixels $(i,\,j)$.
To summarize, the resulting density texture is then computed by
\begin{equation}\label{d0}
d(i,\,j) = d_r(i,\,j) + d_0.
\end{equation}
Although the additive constant value $d_0$ could be any, it should not be too small relative to the maximum density, i.e., $d_0$ should depend on the number of samples $n$ and the total number of pixels $n_p$ in the texture. \edit{}{The average number of samples per pixel $n / n_p$ is the theoretical density of the perfect uniform distribution.} We used \edit{the average number of samples per pixel $n / n_p$ } this constant in our experiments, which always worked well.

\subsection{Smooth Global Deformation}\label{sec:deformation}

Mapping~\ref{map} proposed by Molchanov and Linsen~\cite{Molchanov20_wscg} enlarges the selected region of interest when the density value assigned to the interior part of the selection is larger than the background density. However, when both density values are equal, i.e., the density is constant over the entire visualization domain, mapping~\ref{map} destroys the uniformity of the distribution. Thus, it cannot serve as a density-equalizing transformation.

We fix this issue by computing the defect $t(x,\,y;\,d_0)$, i.e., the distortion mapping of the constant density texture of value  $d_0$, and subtracting it from transformation~\ref{map}. Then, the transformation
\begin{equation}\label{map_res}
    t(x,\,y) = (x,\,y) + t(x,\,y;\,d) - t(x,\,y;\,d_0)
\end{equation}
is an identity mapping for constant density textures: When samples are distributed evenly, $d=d_0$ and $t(x,\,y) = (x,\,y)$. \edit{I.e., iteratively applied deformation has no effect as soon as the samples' density is equalized. }{This adjustment allows for iterative application of the mapping to scatterplot data such that the iterative process converges towards a nearly uniform data distribution. } Convergence of the proposed iterative regularization is demonstrated in our numerical tests in Section~\ref{sec:results}.
Using transformation~\ref{map_res}, all pixels of the texture are mapped to new positions. To compute the new positions of data samples $\mathbf{z}_i$, we perform a bi-linear interpolation using the mappings of the four closest pixels.

\subsection{Algorithm}
\label{sec:algorithm}

The proposed iterative algorithm consists of the following steps:
\begin{enumerate}
    \item For a given set of 2D samples $\mathbf{z}_i$, we generate a smooth rasterized density function representing their distribution in the scatterplot domain. We generate it by summing up per-pixel sample contributions and storing the sum in an accumulation texture. Then, we apply a convolution operator corresponding to a smooth kernel with control parameter $r$, which results in a smooth density plot $d(i,\,j)$.
    \item For the generated discrete density, we compute all ordinary and tilted InIms involved in the computation of mapping~\ref{map_res}. Note that the defect mapping $t(x,\,y;\,d_0)$ in~\ref{map_res} does not depend on $d$. Therefore, $t(x,\,y;\,d_0)$ is computed only once and then reused in each iteration of the algorithm.
    \item Transformation~\ref{map_res} determines images of texture pixels. The per-pixel mapping is approximated at the samples' positions using a bi-linear interpolation formula. New positions of the samples are then computed accordingly.
    \item Steps 1-3 are repeated for the new sample distribution. The stopping criterion can be defined by the user depending on the application domain and analysis task. For instance, the user may want to terminate the regularization process after a fixed number of iterations, after a given elapsed computational time, or based on any measure indicating how regular the achieved distribution is or how significant the impact of the last iteration was.
\end{enumerate}
The presented algorithm has linear complexity with respect to the number of samples. Thus, the primary application domain of the method is the interactive exploration of large datasets.

\subsection{Efficient GPU Computations}\label{sec:GPU}
\label{sec:implementation}

We implemented crucial steps of the proposed algorithm on the GPU. In particular, we developed a novel scheme to efficiently compute (tilted) InIms (Step 2). 

In Step 1 of our algorithm, the subroutine accumulates per-pixel contributions of 2D samples in a high-resolution texture and applies a smoothing procedure. The accumulation is done using vertex and fragment shaders with additive blending. When smoothing is performed using a separable kernel, e.g., the Gaussian kernel used in our experiments, the convolution can be performed with two compute-shader passes (vertical and horizontal directions separately).

In Step 2, we compute the InIms involved in the constructions of the mapping.
Singhal et al.~\cite{Singhal12} described computation of \mbox{InIms} on the GPU using a 2D reduction technique performed in the fragment shader. We adopt this algorithm for efficient computation of InIms and extend it to tilted InIms, see Figure~\ref{fig:GPU}. (i) For a given scalar-valued density texture, we first compute upper- (red) and lower-\textit{column integrals} (green) as shown in Figures~\ref{fig:vertcol_1}--\ref{fig:vertcol_3}. The computations are progressive~\cite{Stolper14}, where the summed area doubles in each step. For a square texture of size $2^k\times 2^k$, the procedure requires one rendering pass with $k$ summation operations to complete. \edit{}{Our implementation launches one workgroup for each row/column/diagonal in the image. Each thread only relies on data computed within the same workgroup. As such, we only need a single rendering pass with a for-loop of $k$ iterations and a barrier synchronizing the work group after each iteration. } The resulting two-channel texture contains sums of density values at the pixels located above and below the considered pixel, respectively. (ii) The \textit{classical InIms} can then be found in an analogous multi-pass procedure by collecting the computed column-integral values horizontally as shown in Figures~\ref{fig:inim_1}--\ref{fig:inim_3}. (iii)~For the computation of tilted InIms, we precompute four \textit{auxiliary triangle integrals} as presented in Figures~\ref{fig:triag_1}--\ref{fig:triag_3}. Their calculation is based on summing up the column integrals along respective diagonals. (iv) Finally, the \textit{tilted InIms} can be easily computed by performing arithmetic operations on the auxiliary triangle textures, see Figures~\ref{fig:tilt_1}--\ref{fig:tilt_3} for $\alpha_t$. In the figure, the upper-left and upper-right auxiliary triangle integrals are summed and the twice-counted column integral is subtracted. The other three tilted integrals $\alpha_t$, $\beta_t$, and $\gamma_t$ can be computed analogously. 

In Step 3, the two-channel deformation texture $t(x,\,y)$ can then be calculated according to~\ref{map_res} at every pixel. Bi-linear interpolation of the deformation texture $t(x,\,y)$ at the sample positions results in a new set of mapped 2D samples.
The dependence of the execution time on the algorithm's parameters and the data size is extensively explored in the numerical tests presented in Section~\ref{sec:results}.

\newcommand{\mysizeS}{0.25}
\begin{figure}[!tb]
\centering
\subfloat[]{\includegraphics[width=\mysizeS\linewidth]{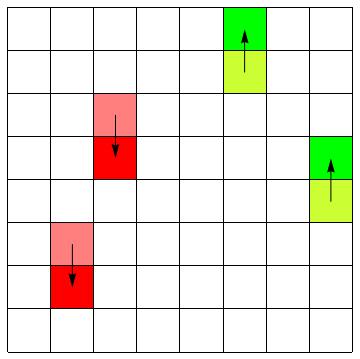}\label{fig:vertcol_1}}
\raisebox{\dimexpr 0.9cm}{\hphantom{$\mathbf{-}$}}
\subfloat[]{\includegraphics[width=\mysizeS\linewidth]{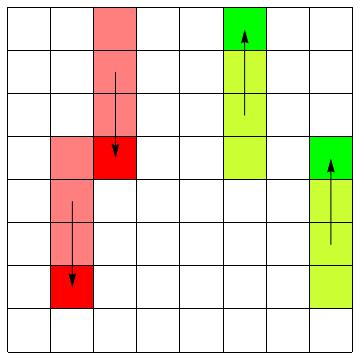}\label{fig:vertcol_2}}
\raisebox{\dimexpr 0.9cm}{\hphantom{$\mathbf{-}$}}
\subfloat[]{\includegraphics[width=\mysizeS\linewidth]{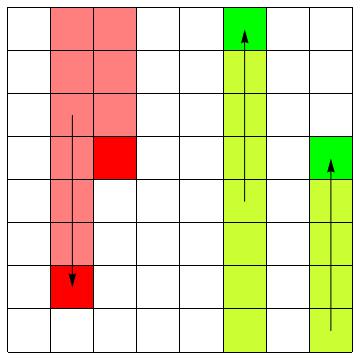}\label{fig:vertcol_3}}\\
\subfloat[]{\includegraphics[width=\mysizeS\linewidth]{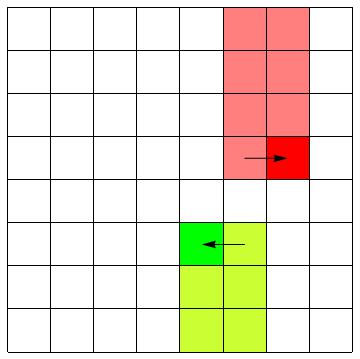}\label{fig:inim_1}}
\raisebox{\dimexpr 0.9cm}{\hphantom{$\mathbf{-}$}}
\subfloat[]{\includegraphics[width=\mysizeS\linewidth]{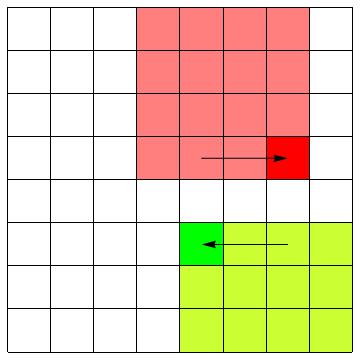}\label{fig:inim_2}}
\raisebox{\dimexpr 0.9cm}{\hphantom{$\mathbf{-}$}}
\subfloat[]{\includegraphics[width=\mysizeS\linewidth]{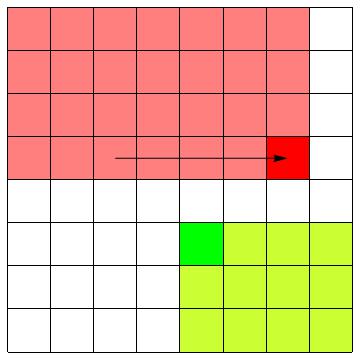}\label{fig:inim_3}}\\
\subfloat[]{\includegraphics[width=\mysizeS\linewidth]{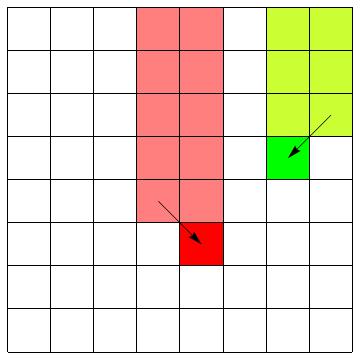}\label{fig:triag_1}}
\raisebox{\dimexpr 0.9cm}{\hphantom{$\mathbf{-}$}}
\subfloat[]{\includegraphics[width=\mysizeS\linewidth]{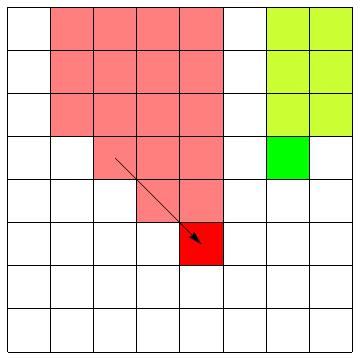}\label{fig:triag_2}}
\raisebox{\dimexpr 0.9cm}{\hphantom{$\mathbf{-}$}}
\subfloat[]{\includegraphics[width=\mysizeS\linewidth]{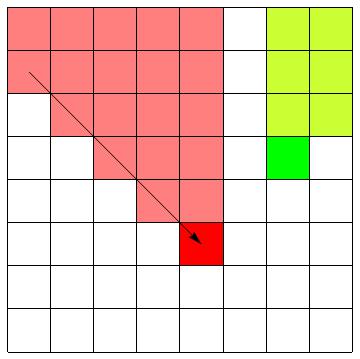}\label{fig:triag_3}}\\
\subfloat[]{\includegraphics[width=\mysizeS\linewidth]{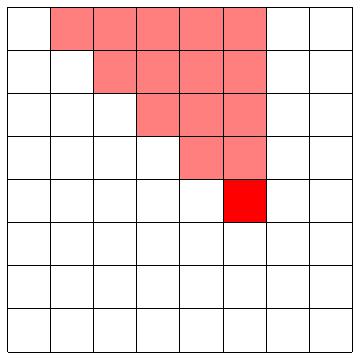}\label{fig:tilt_1}}
\raisebox{\dimexpr 0.9cm}{\mbox{\Huge$\mathbf{+}$}}
\subfloat[]{\includegraphics[width=\mysizeS\linewidth]{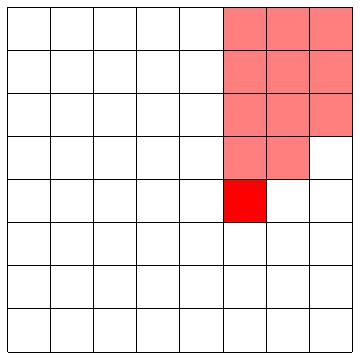}\label{fig:tilt_2}}
\raisebox{\dimexpr 0.9cm}{\mbox{\Huge$\mathbf{-}$}}
\subfloat[]{\includegraphics[width=\mysizeS\linewidth]{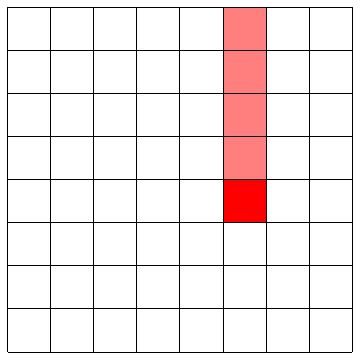}\label{fig:tilt_3}}
\caption{Efficient computation of InIms.
\protect\subref{fig:vertcol_1}--\protect\subref{fig:vertcol_3}~Computation of column integrals. Pixels highlighted in red iteratively accumulate texture values from the pixels located above them. Green pixels progressively sum up values from the pixel columns below. After $k$ summations are performed in a single rendering pass, every pixel contains upper- and lower-column sums of values stored in two texture channels.
\protect\subref{fig:inim_1}--\protect\subref{fig:inim_3}~Computation of InIms by iterative accumulation of column integrals. In the last iteration, red and green pixels contain values for $\alpha$ and $\gamma$, correspondingly. InIms for $\beta$ and $\delta$ can be computed analogously. Note that $\beta$, $\gamma$, and $\delta$ can be evaluated on demand using $\alpha$ only, thus, their explicit computation is not necessary.
\protect\subref{fig:triag_1}--\protect\subref{fig:triag_3}~Calculation of triangle integrals by summing up column integrals along diagonals. Two of the four required auxiliary integrals are shown.
\protect\subref{fig:tilt_1}--\protect\subref{fig:tilt_3}~Tilted InIms can be computed by simple arithmetic operations on precomputed column and triangle integrals. An example for calculating $\alpha_t$ is presented. Tilted InIms $\beta_t$, $\gamma_t$, and $\delta_t$ can be found analogously.
}
\label{fig:GPU}
\end{figure}

\section{Visual Encoding of Local Deformation}\label{sec:preservation}

Structures and patterns (such as clusters, outliers, dense and sparse regions) of the samples' distribution in scatterplots may reveal features and characteristics of the studied phenomena. Occlusion hinders managing and accessing the samples when the data size is large. Our approach mitigates the overplotting issue by regularizing the samples' distribution in the scatterplot. A regularized layout is by definition patternless. Though the regularized layout is not designed to completely replace the original configuration, preserving the main data features of the original scatterplots could be beneficial. Recognition of patterns of the original scatterplot after deformation motivates us to explore visual encodings, which could help to convey data structures, at least partially. The approaches discussed in this section are evaluated in the user study of Section~\ref{sec:user_study} (cf. Task~T3). 

Our mapping~\ref{map_res} is an affine combination of points on the domain boundary with non-negative weights that vary smoothly. Moreover, since the InIms are by construction monotonic, i.e., they have no extreme points in the interior of the domain, the resulting mapping also preserves some degree of monotonicity. In particular, it does not introduce any discontinuities or twisting/swirling behavior. Thus, neighborhood relations of samples are generally preserved, meaning that points in the neighborhood of a point will remain in the neighborhood after deformation. However, this does not mean that the $k$ nearest neighbors will remain the $k$ nearest neighbors, as distances between points will, in general, be distorted anisotropically. 
Preservation of neighborhoods is important since the regularized layout facilitates accessing and managing samples. Preservation of other characteristics of the undistorted distribution is per se not the goal of the de-cluttering approach, but could nevertheless be beneficial for the user.
Indeed, since distances between samples are relaxed, some data analysis tasks cannot be reliably performed on the regularized \edit{}{plot}. However, additional visual cues may make data analysis possible even without switching back to the original scatterplot layout.

Identifying data clusters and outliers in the distorted map requires the analyst to take into account the local characteristics of the applied transformation. Such information needs to be visually conveyed. 
We propose to use different visual encodings of the local deformations and evaluate them in a user study.

As a first option, we follow the approach by Carpendale et al.~\cite{Carpendale95} and draw a \emph{sparse regular grid}, whose nodes we deform just like the data samples by using transformation~\ref{map_res}. Based on the local density of the grid lines after deformation, one can judge the local area expansion or contraction factor as well as its direction. The method is simple and has been applied successfully in other studies. However, it introduces additional geometry, which increases visual complexity and thus adds to the occlusion in the plot. 

As a second option, we propose to use a \emph{density background texture}. The density field of the given scatterplot is computed as described in Section~\ref{sec:densityfield}. It is then deformed using our mesh mapping from Section~\ref{sec:deformation}.
Finally, we apply a transfer function to map densities to colors and render the resulting texture as the background of our deformed scatterplot. Consequently, we expect high-density areas such as data clusters to remain visible after distortion. More precisely, high-intensity regions denote the cores of the clusters and the low-intensity regions depict the separation of clusters.
Alternatively, when the background texture is used for other purposes,
one may use a \emph{density color-coding of the samples}. Thus, one encodes the original density at the positions of the transformed samples using color. This option follows the ideas that are also used in appearance change for clutter reduction, see Section~\ref{sec:related_work}. We therefore leave this alternative out of the scope of this paper.

\renewcommand{\mysize}{0.47}
\newcommand{\mysizetwo}{0.49}
\begin{figure}[!htb]
  \centering
  \subfloat[\editMinor{original layout}{}]{\includegraphics[width=\mysize\linewidth]{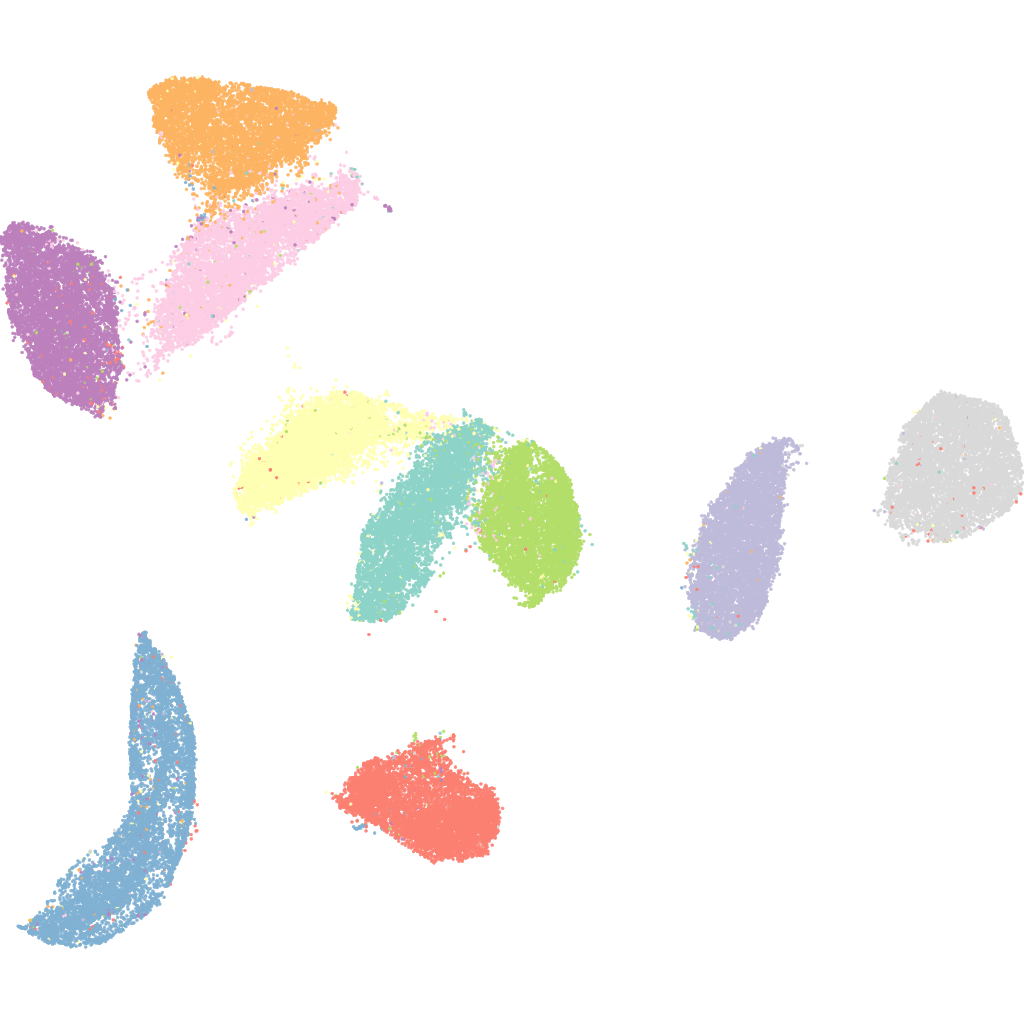} \label{fig::mnist_orig}}\ 
  \subfloat[\editMinor{de-cluttered with grid lines}{}]{\includegraphics[width=\mysize\linewidth]{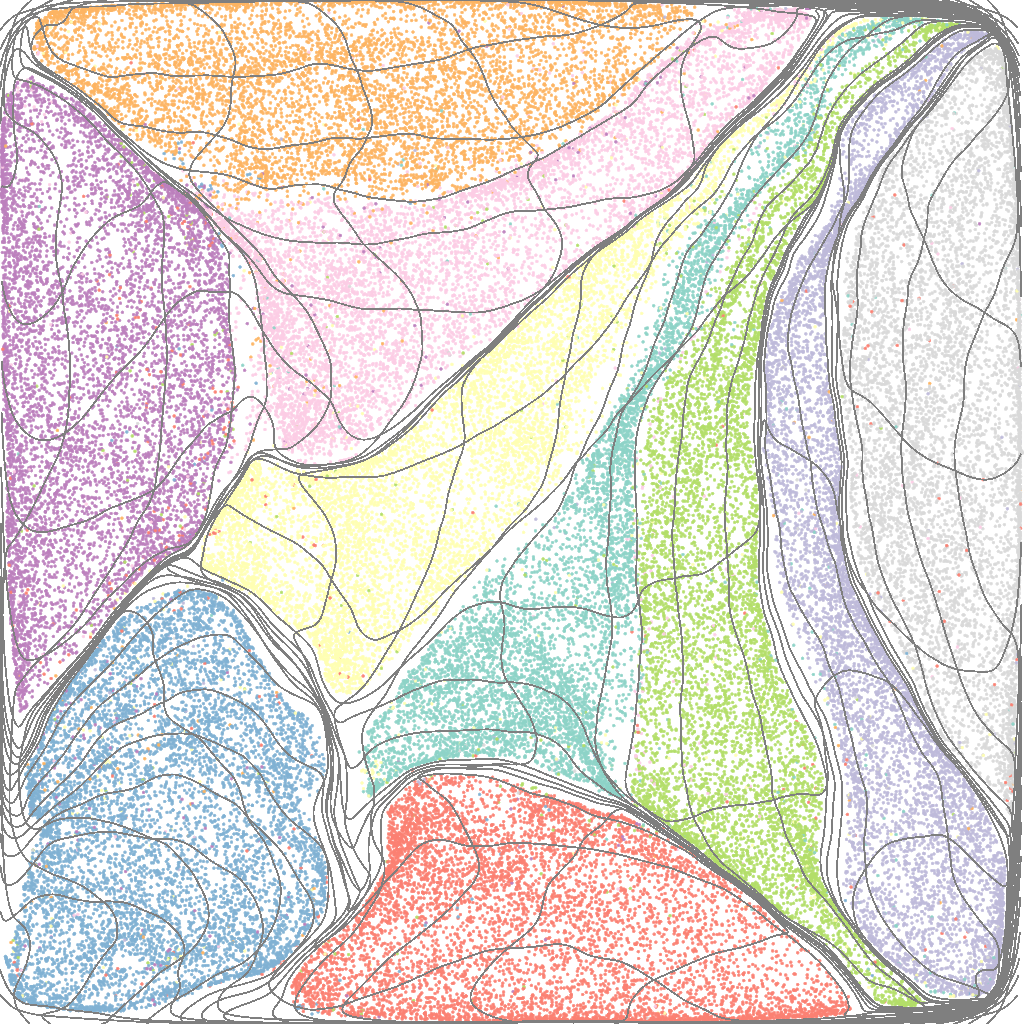} \label{fig::mnist_grid}}\\
  \subfloat[\editMinor{de-cluttered with density texture}{}]{\includegraphics[width=\mysize\linewidth]{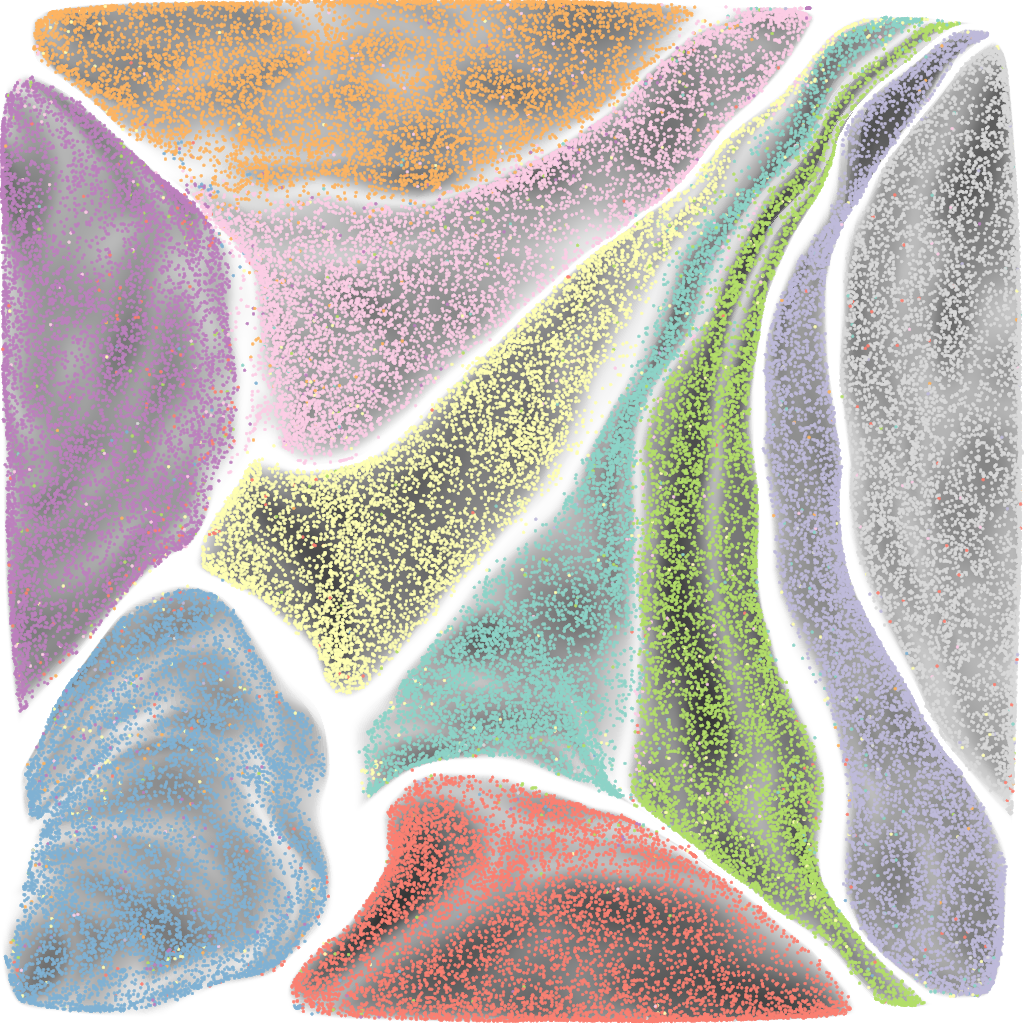} \label{fig::mnist_density}}\ 
  \subfloat[\editMinor{de-cluttered with contour lines}{}]{\includegraphics[width=\mysize\linewidth]{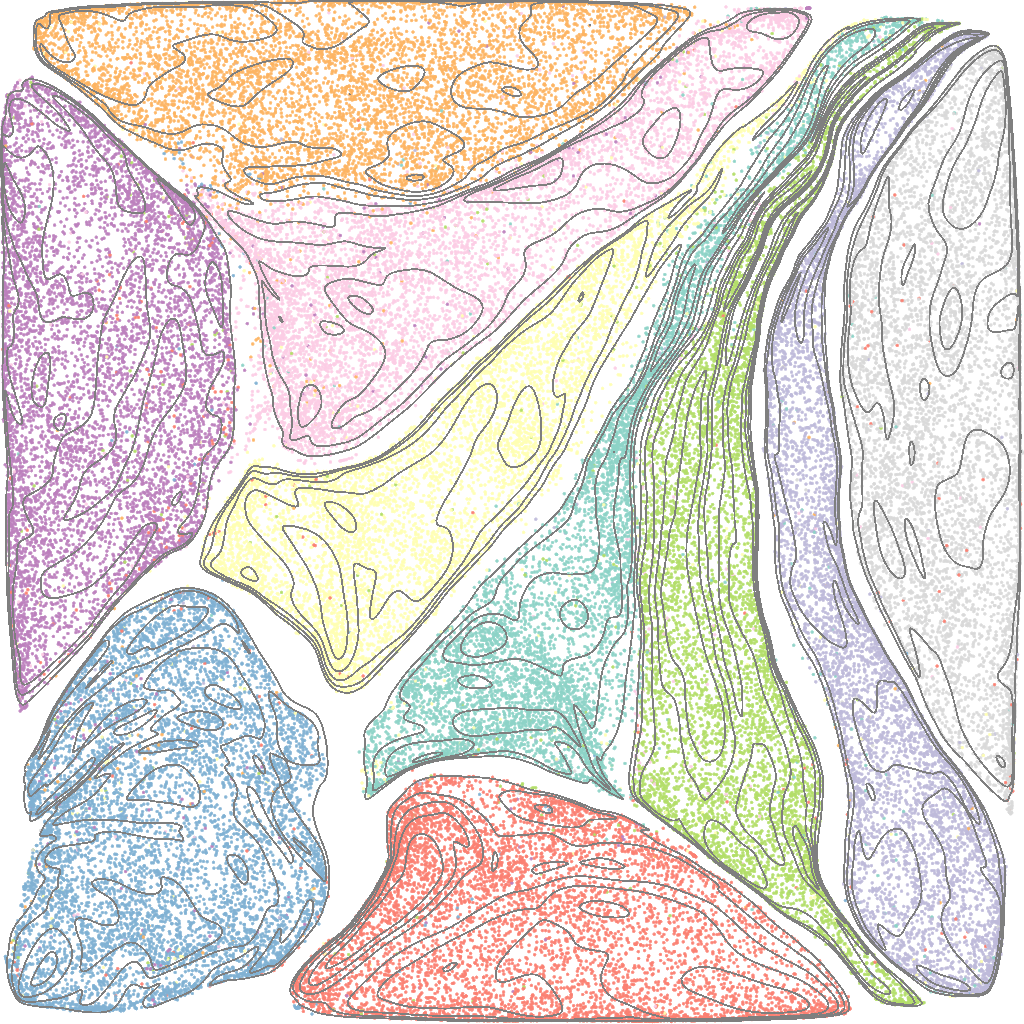} \label{fig::mnist_contour}}
\caption{\protect\subref{fig::mnist_orig} Original layout of the MNIST dataset (UMAP, number of neighbors~$15$, minimal distance~$0.1$) with color-coded classes. \protect\subref{fig::mnist_grid} Visual encoding of the density-equalizing transform using grid lines after~$32$ iterations. The original density of samples is represented by the background texture in~\protect\subref{fig::mnist_density} and by contour lines in~\protect\subref{fig::mnist_contour}. The last two figures allow for analyzing the subcluster structures occluded in~\protect\subref{fig::mnist_orig}.
}
\label{fig::mnist}
\end{figure}

Finally, density variation within original data clusters can be roughly depicted using contour lines of the density function. Showing the same contour lines after density-equalizing transformation allows for identifying the clusters' boundaries. Figure~\ref{fig::mnist} compares visual encodings of the applied transform using grid lines, a density texture, and contour lines after regularization of a UMAP embedding of the MNIST dataset~\cite{Lecun98}. Figure~\ref{fig::mnist_density} and Figure~\ref{fig::mnist_contour} reveal subcluster structures that could not be inspected in the original scatterplot due to high occlusion. Results of the user study comparing the benefits and effectiveness of different approaches are presented in Section~\ref{sec:user_study}.

\section{Results}\label{sec:results}

We first present some results that visually show the de-cluttering process on various scatterplot examples (Section~\ref{sec:visual_results}), then perform some numerical tests on our algorithm in terms of performance, quality, and stability (Section~\ref{sec:numerical_tests}), and finally describe the user study we conducted (Section~\ref{sec:user_study}).

\subsection{Visual Investigations}
\label{sec:visual_results}

In the first experiment, we illustrate the regularization effect of the proposed algorithm. We performed a few iterations starting with samples distributed roughly along the domain diagonal (Figure~\ref{fig::func_line_0}).
The deformation mapping is conveyed by showing a transformation of a regular grid throughout the iterations. When samples are placed along the diagonal, distributions of the data in $x$ and $y$ dimensions are uniform. Therefore, the application of the HistoScale-based approach proposed by Keim et al.~\cite{Keim03} would have no effect. Our algorithm efficiently deforms the visual domain spreading the samples from the diagonal over the available void space.

\renewcommand{\mysize}{0.31}
\renewcommand{\mysizetwo}{0.49}
\begin{figure}[!t]
  \centering
  \includegraphics[width=\mysize\linewidth]{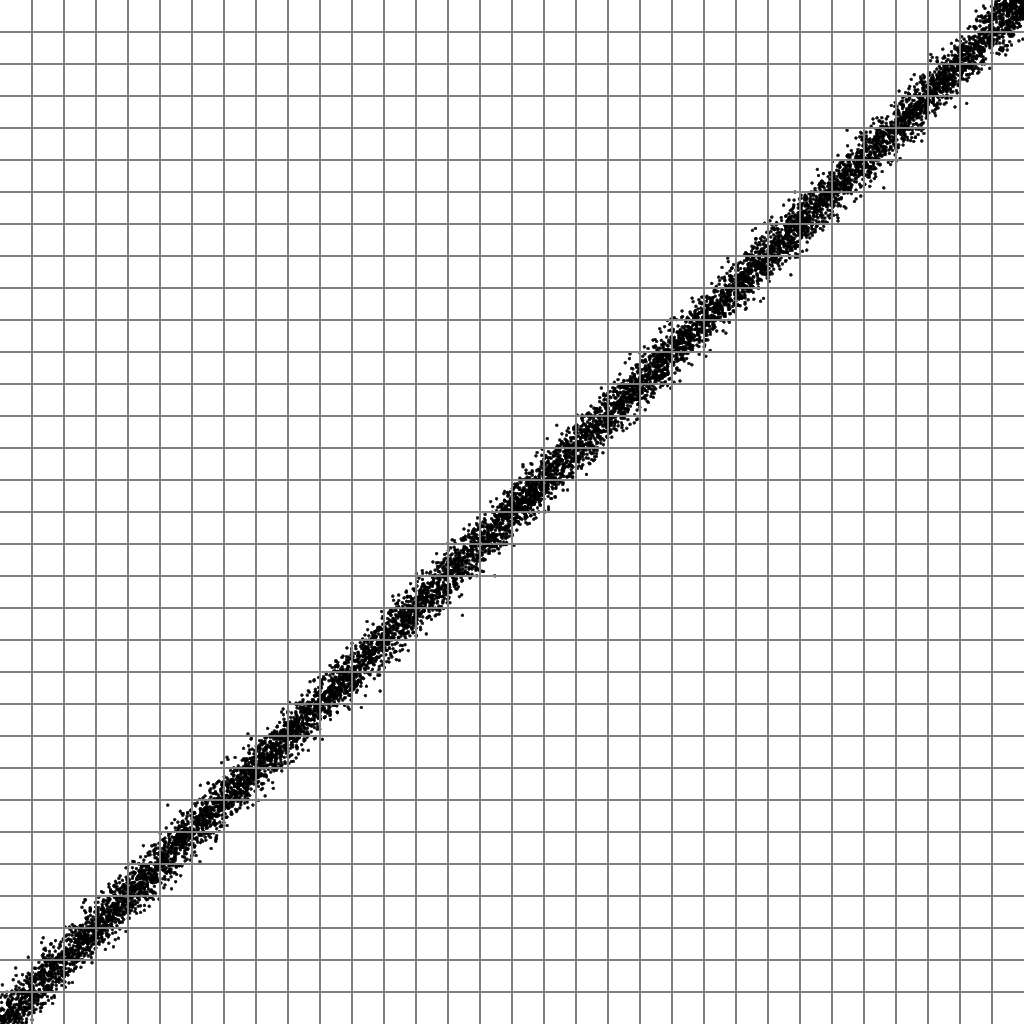} \label{fig::func_line_0}\ 
  \includegraphics[width=\mysize\linewidth]{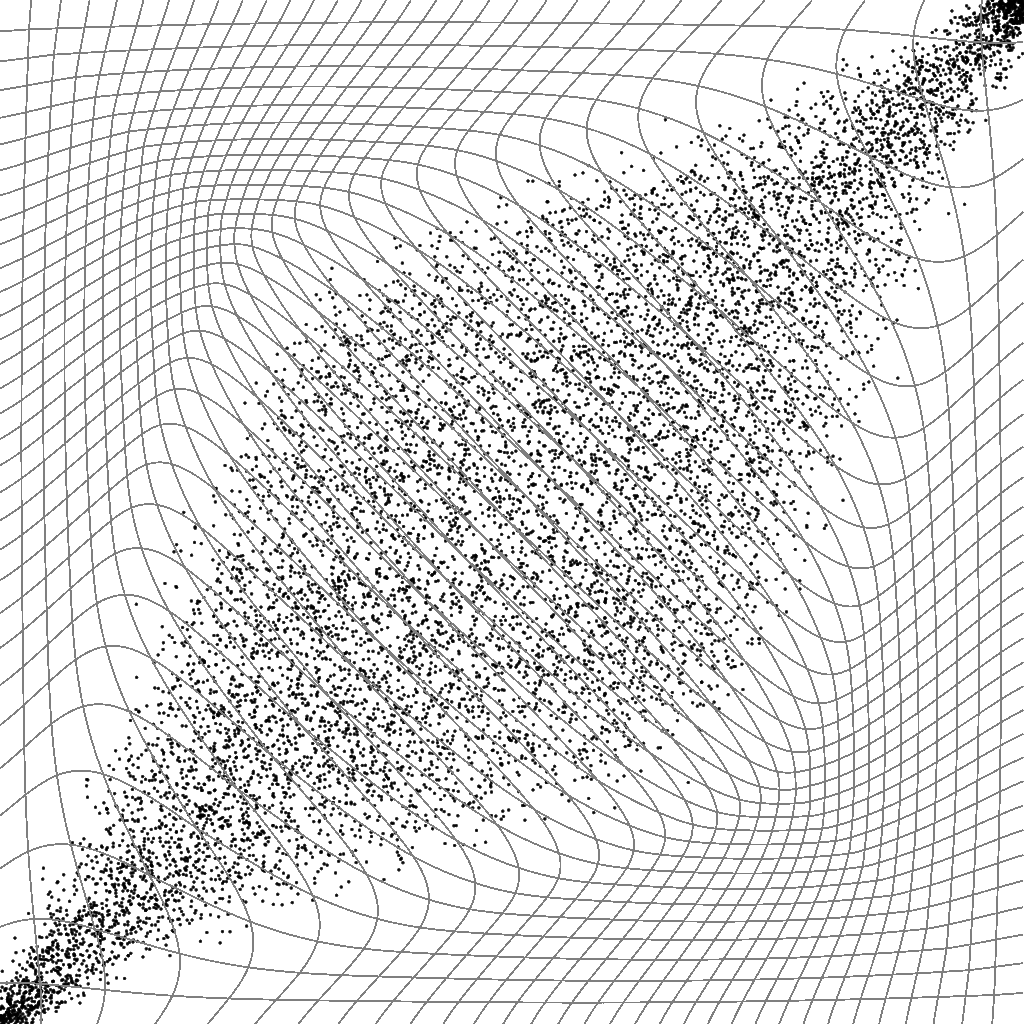} \label{fig::func_line_2}\ 
  \includegraphics[width=\mysize\linewidth]{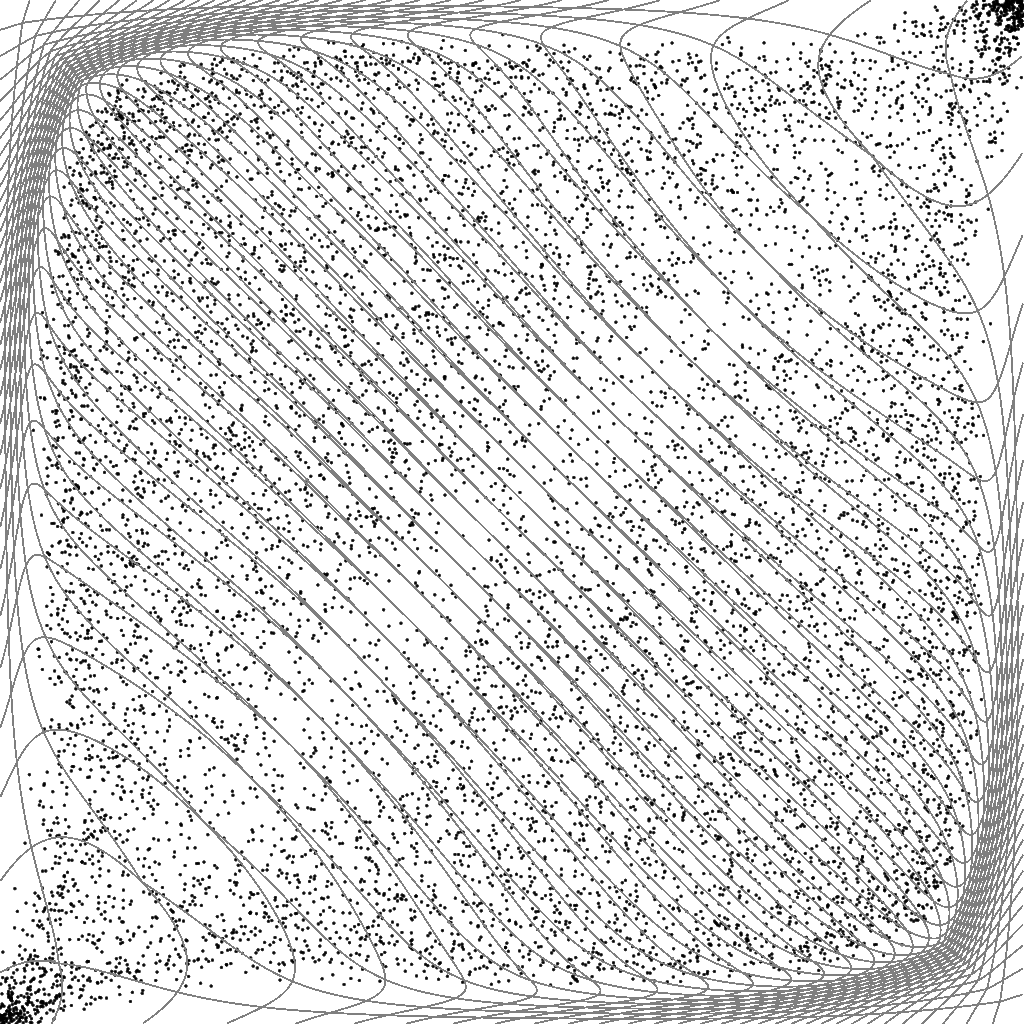} \label{fig::func_line_8}
\caption{Regularization of data sampled roughly along the domain diagonal. A superimposed regular grid conveys the domain deformation. \editMinor{}{Left: Original scatterplot. Middle: After 2 iterations. Right: After 8 iterations.}}
\label{fig::func_diag}
\end{figure}

In a second experiment, we demonstrate that no mixing of samples from different clusters takes place during the deformation presented in Figure~\ref{fig::reg}.
During iterative transformations, the four clusters significantly change their shape such that the free space between the clusters is contracted. Still, due to the smooth and monotonic character of the transformation, no samples from one cluster can be mapped to the area occupied by the samples from the other cluster. All clusters remain separated by thin unoccupied areas distinguished by a high density of grid lines. 
Figure~\ref{fig::mix} shows a data set with three clusters, where within each cluster regions of different shapes are manually selected in the \edit{origical }{original } scatterplot (selections are highlighted by color).  We observe that the selection boundaries remain sharp after regularization. Thus, even closely related samples within the same cluster preserve their local order. Yet another experiment demonstrating the preservation of the samples' neighborhoods is shown in the accompanying video. 

\renewcommand{\mysize}{0.29}
\renewcommand{\mysizetwo}{0.49}
\begin{figure}[!t]
  \centering
  \subfloat[\editMinor{original scatterplot}{}]{\includegraphics[width=\mysize\linewidth]{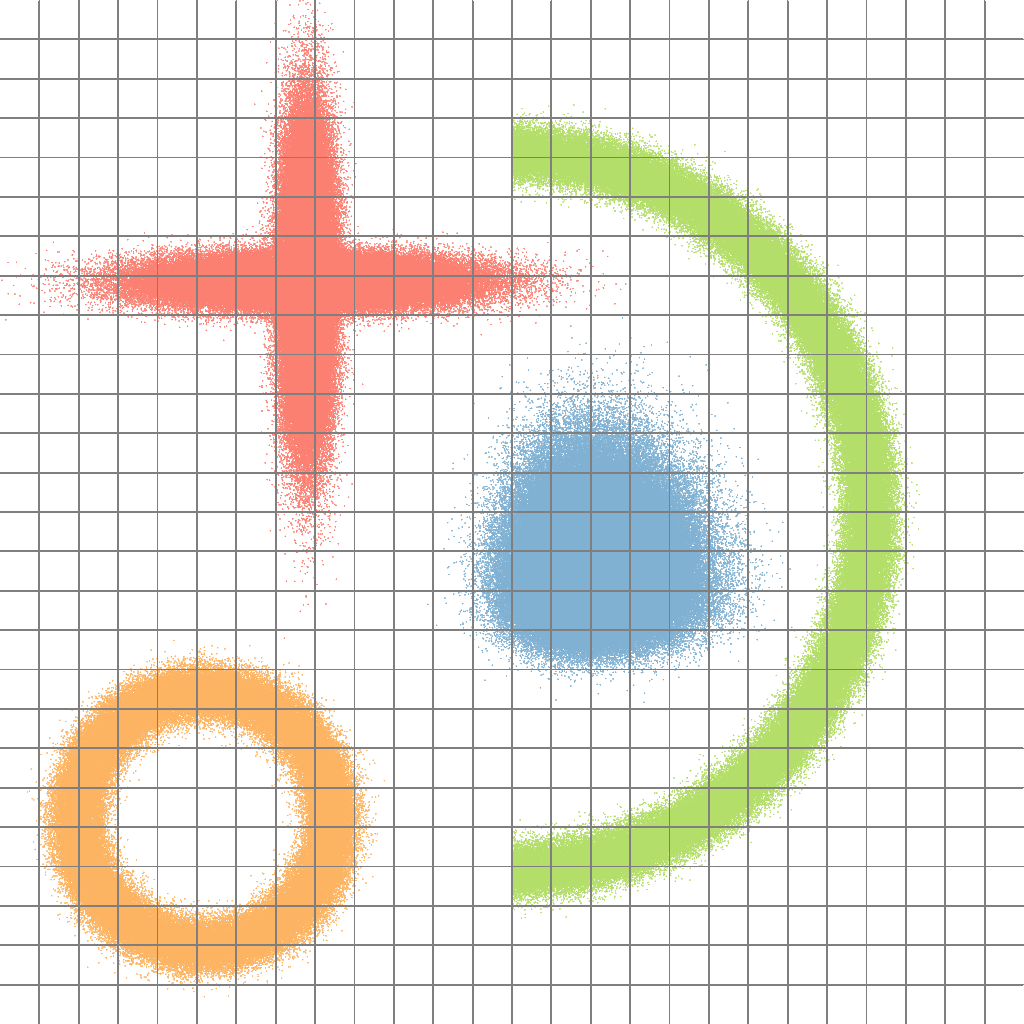} \label{fig::reg_0}}\quad
  \subfloat[\editMinor{iter. 1, $1.15$~ms}{}]
  {\includegraphics[width=\mysize\linewidth]{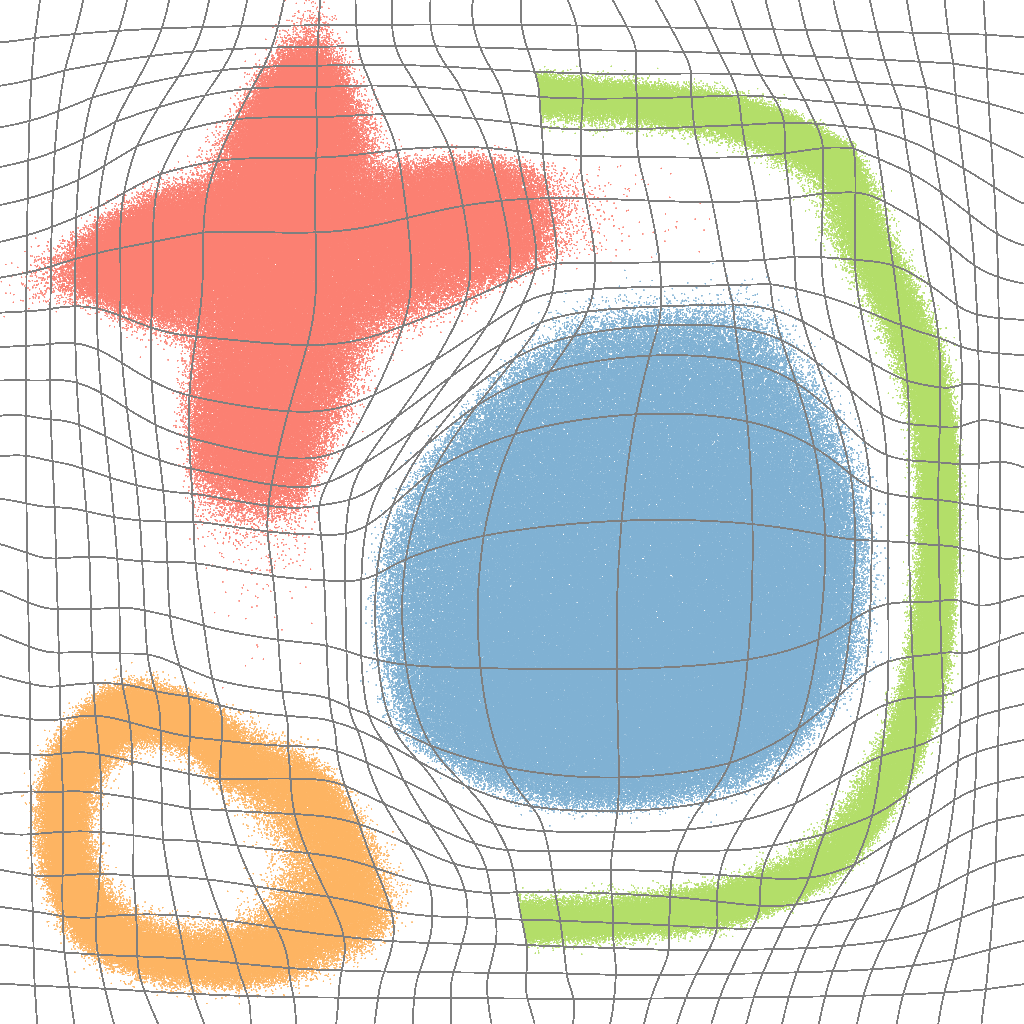} \label{fig::reg_1}}\quad
  \subfloat[\editMinor{iter. 2, $2.99$~ms}{}]
  {\includegraphics[width=\mysize\linewidth]{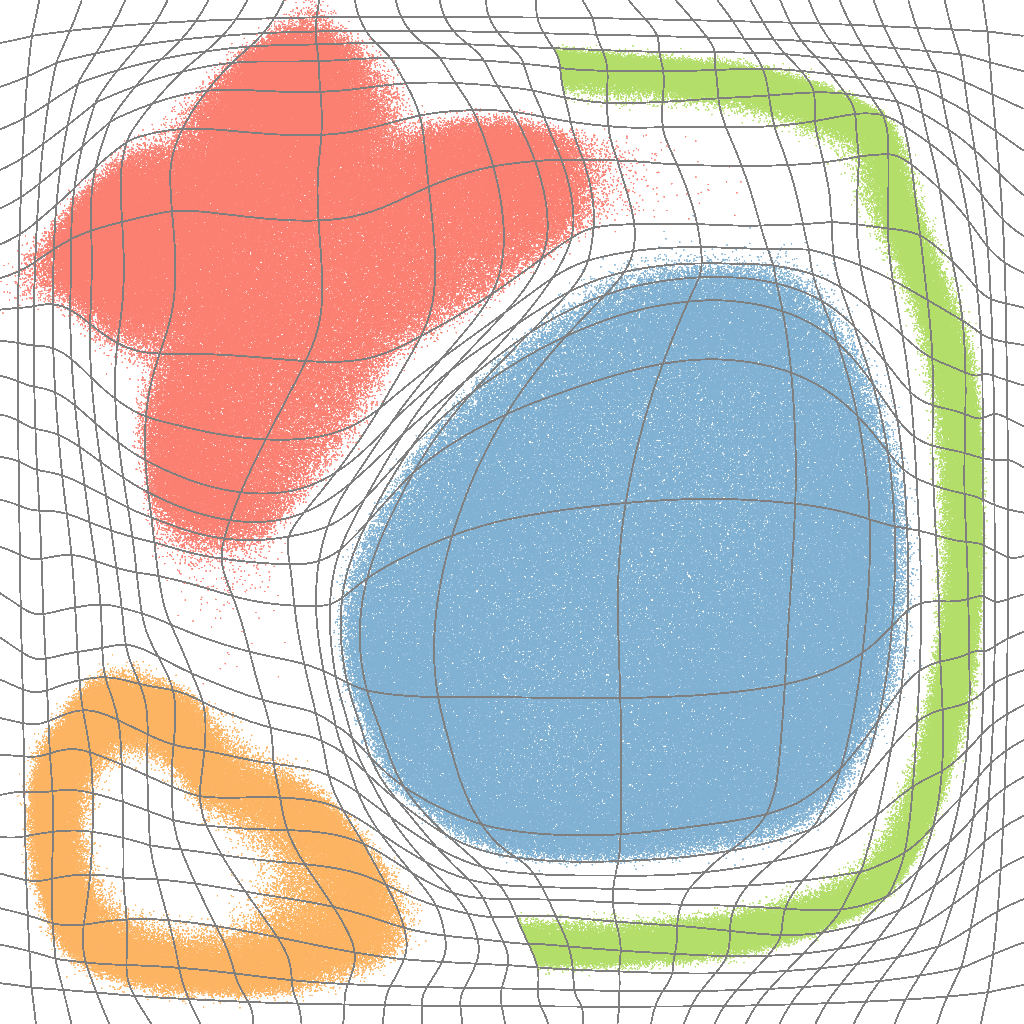} \label{fig::reg_2}}\\
    \subfloat[\editMinor{iter. 4, $6.22$~ms}{}]
    {\includegraphics[width=\mysize\linewidth]{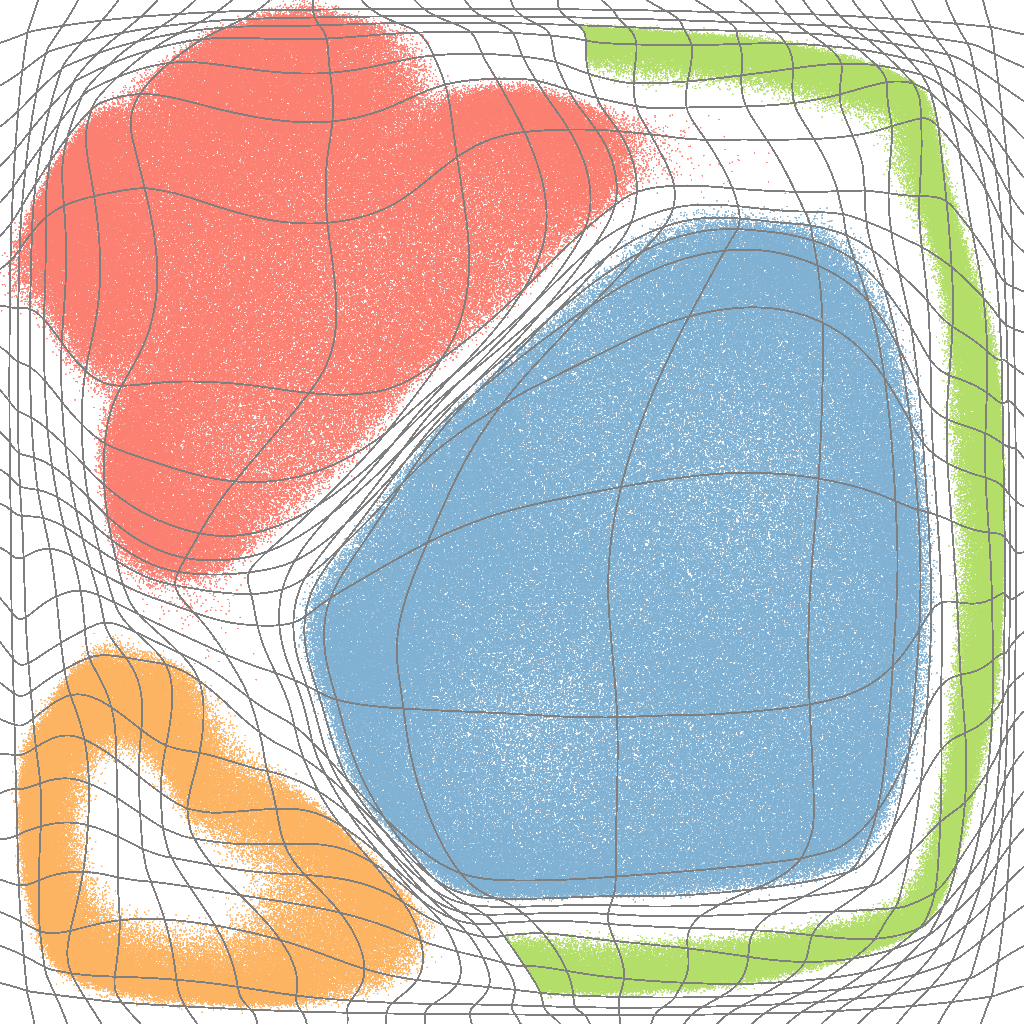} \label{fig::reg_4}}\quad
  \subfloat[\editMinor{iter. 8, $12.76$~ms}{}]
  {\includegraphics[width=\mysize\linewidth]{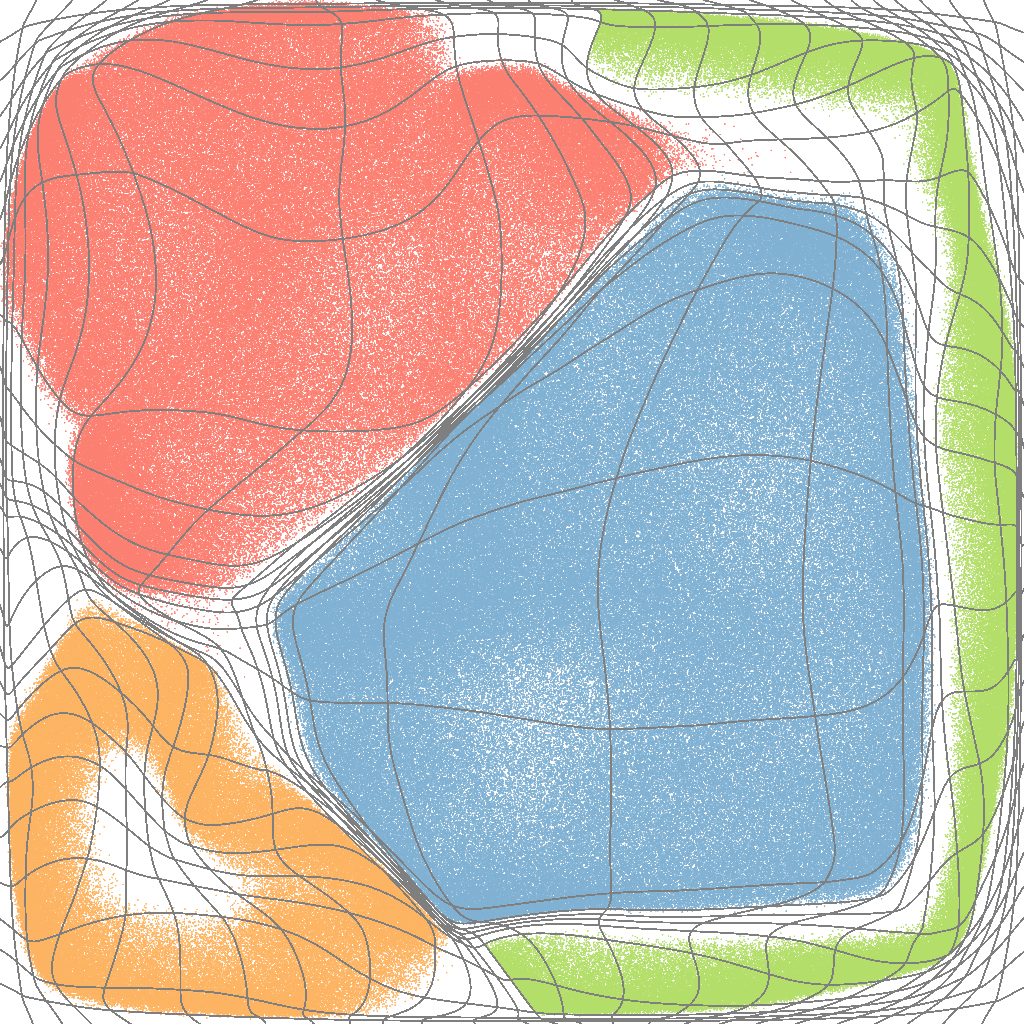} \label{fig::reg_8}}\quad
  \subfloat[\editMinor{iter. 16, $25.24$~ms}{}]
  {\includegraphics[width=\mysize\linewidth]{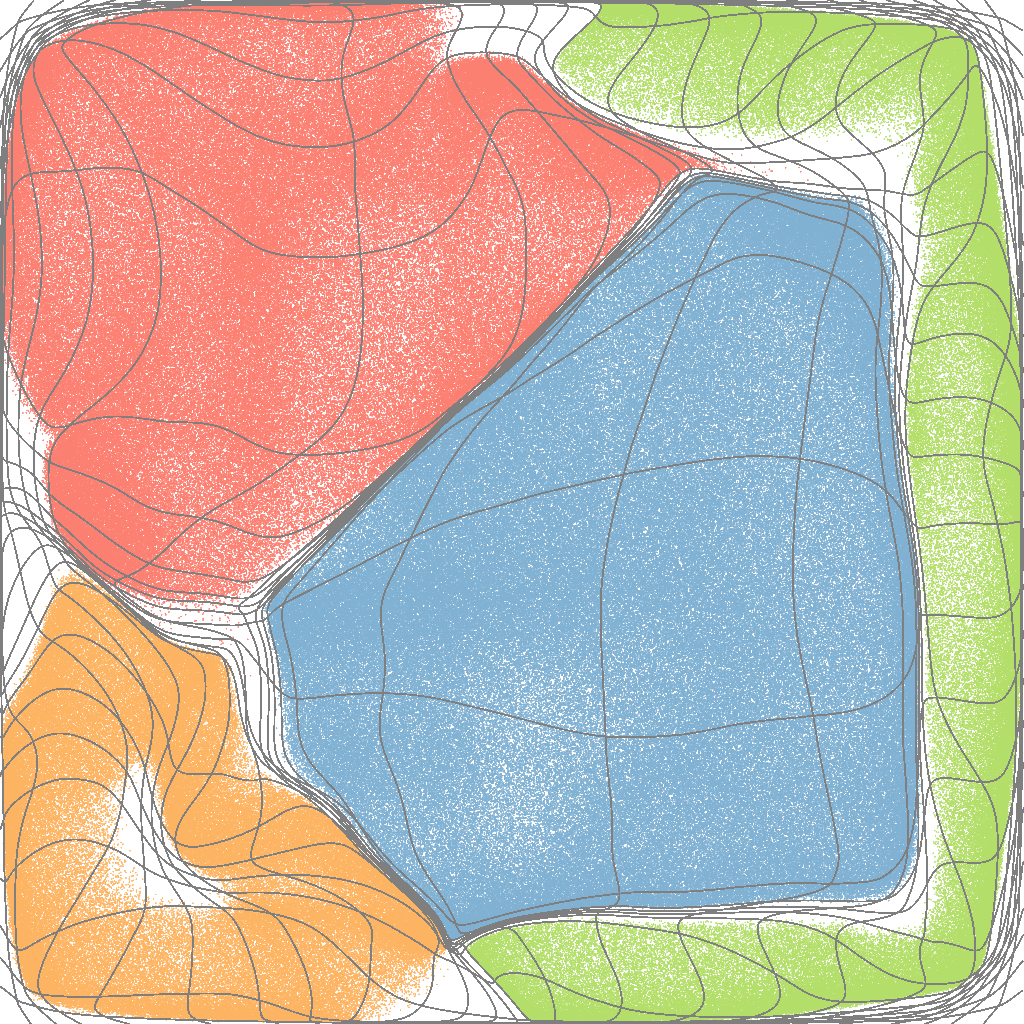} \label{fig::reg_16}}\\
\subfloat[\editMinor{Generalized Scatterplot,  distortion~$1$, overlap~$0.1$}{}]
{\includegraphics[width=\mysize\linewidth]{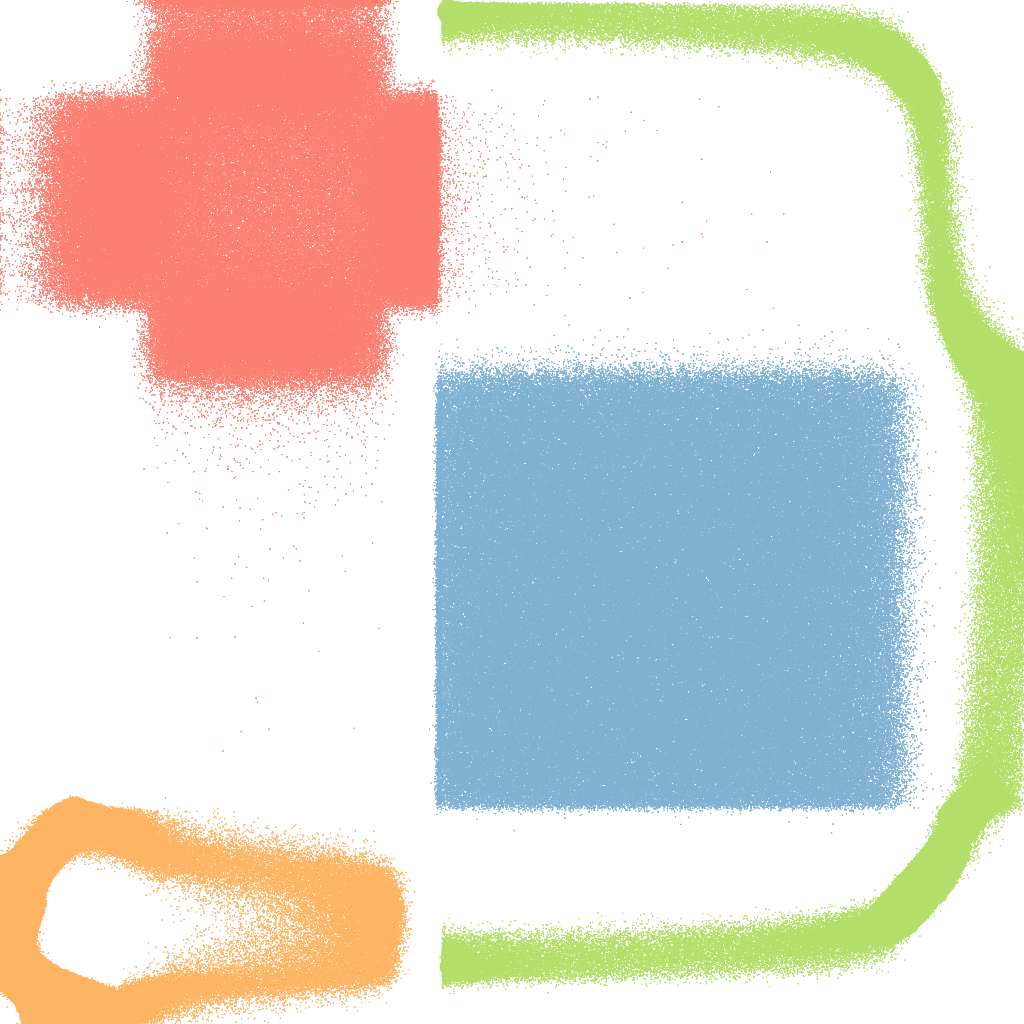} \label{fig::keim_0}}\quad
  \subfloat[\editMinor{Generalized Scatterplot, distortion~$0.5$, overlap~$0.05$}{}]
  {\includegraphics[width=\mysize\linewidth]{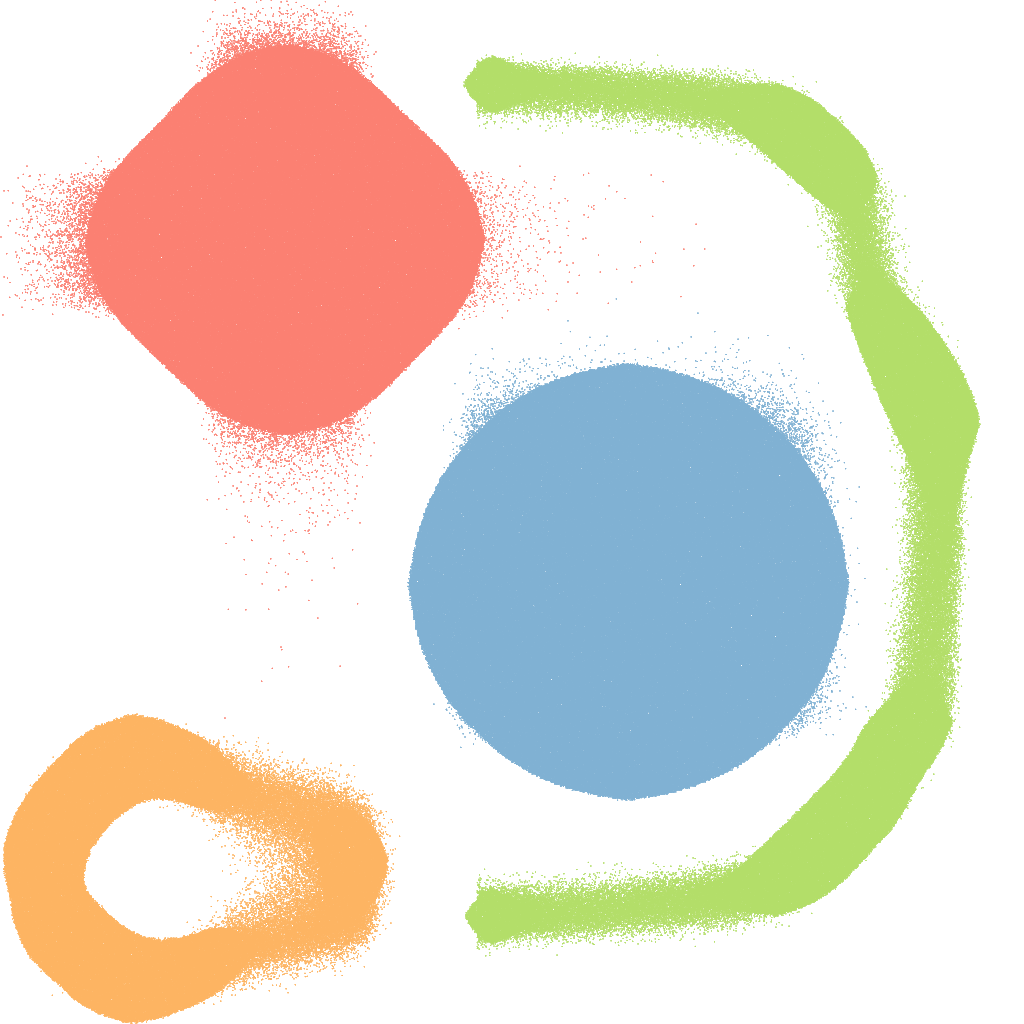} \label{fig::keim_1}}\quad
  \subfloat[\editMinor{Generalized Scatterplot, distortion~$0$, overlap~$0.1$}{}] {\includegraphics[width=\mysize\linewidth]{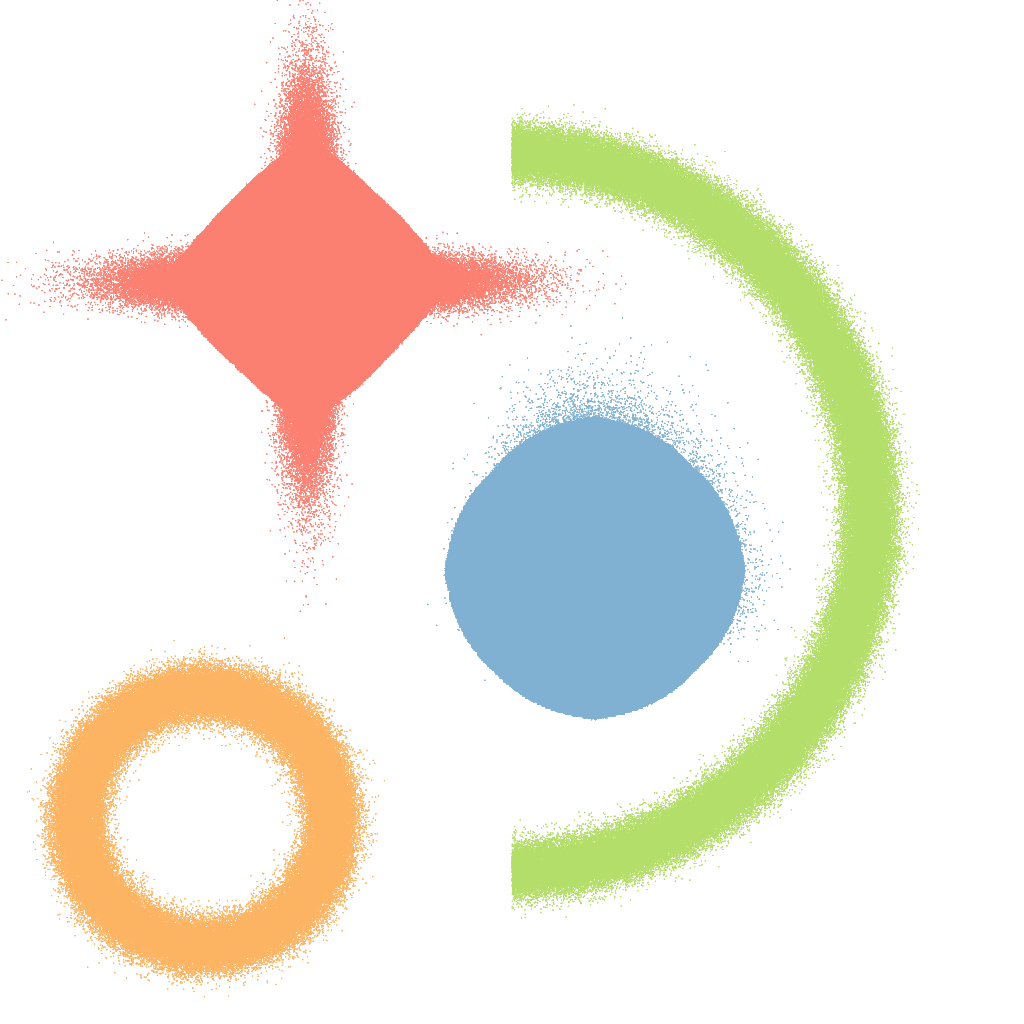}\label{fig::keim_2}}
\caption{Regularization of samples' distribution in scatterplot. \protect\subref{fig::reg_0}~Original scatterplot depicts four clusters shown in blue (400k samples), red (300k samples), green (200k samples), and \edit{yellow }{orange } (100k samples). Visual estimation of cluster sizes as well as access to individual samples are hindered by excessive overplotting. \protect\subref{fig::reg_1}-\protect\subref{fig::reg_16} Iterative transformation of scatterplots using the proposed de-cluttering algorithm \editMinor{}{after 1 (b), 2 (c), 4 (d), 8 (e), and 16 (f) iterations. Computational times are $1.15$~ms, $2.99$~ms, $6.22$~ms, $12.76$~ms, and $25.24$~ms respectively}. \editMinor{Computational times are provided. }{}After a few iterations, data clusters occupy areas proportional to the number of samples contained in them. No mixing of clusters takes place. A superimposed regular grid is deformed using the same mapping. The shape of the deformed grid represents the computed mapping and may serve for the identification of the original data clusters even if they were not color-coded. \protect\subref{fig::keim_0}-\protect\subref{fig::keim_2} Generalized Scatterplots proposed by Keim et al.~\cite{Keim10} demonstrate noticeably less efficient use of the screen space for any combination of the governing parameters\editMinor{.}{: (g) distortion~$=1$, overlap~$=0.1$, (h) distortion~$=0.5$, overlap~$=0.05$, (i) distortion~$=0$, overlap~$=0.1$.}}
\label{fig::reg}
\end{figure}

\renewcommand{\mysize}{0.31}
\begin{figure}[!hbt]
  \centering
    \includegraphics[width=\mysize\linewidth]{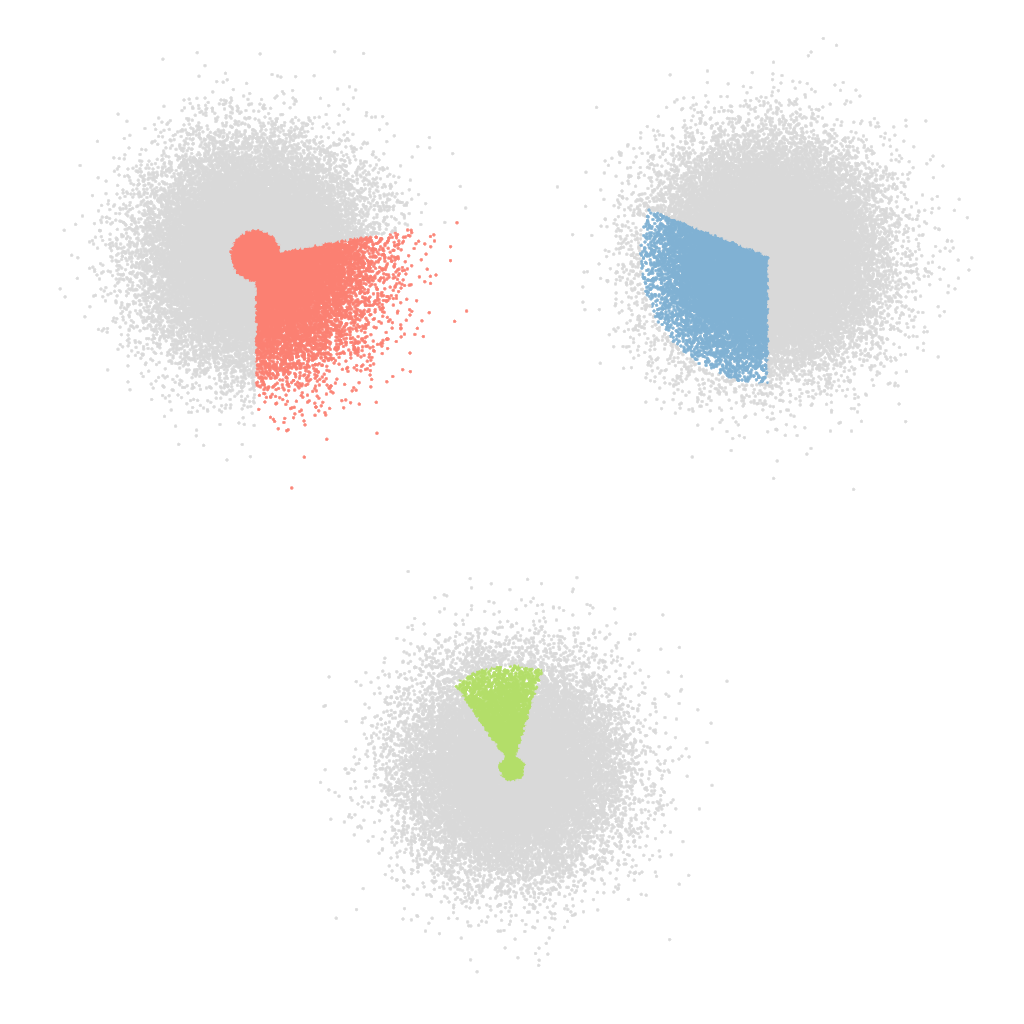}\label{fig::mix_0}\ 
    \includegraphics[width=\mysize\linewidth]{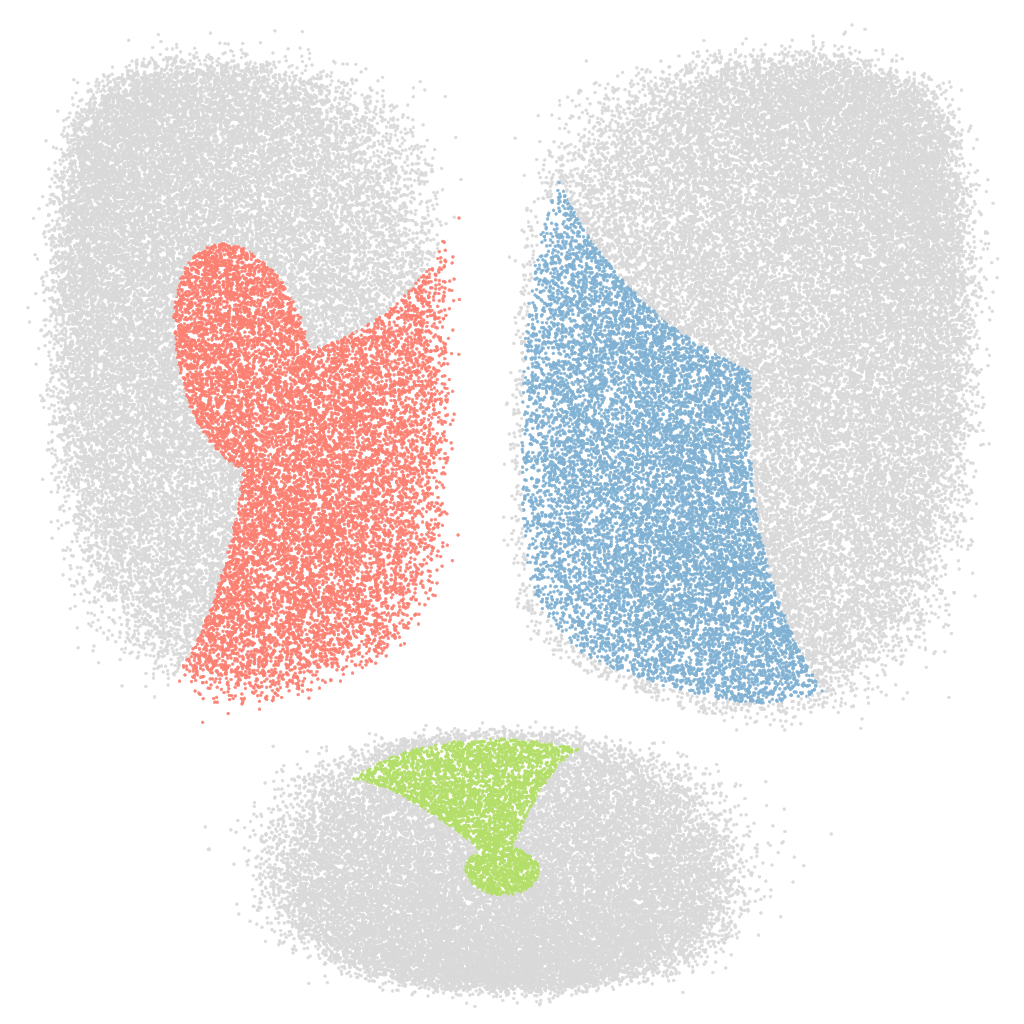}\label{fig::mix_2}\ 
    \includegraphics[width=\mysize\linewidth]{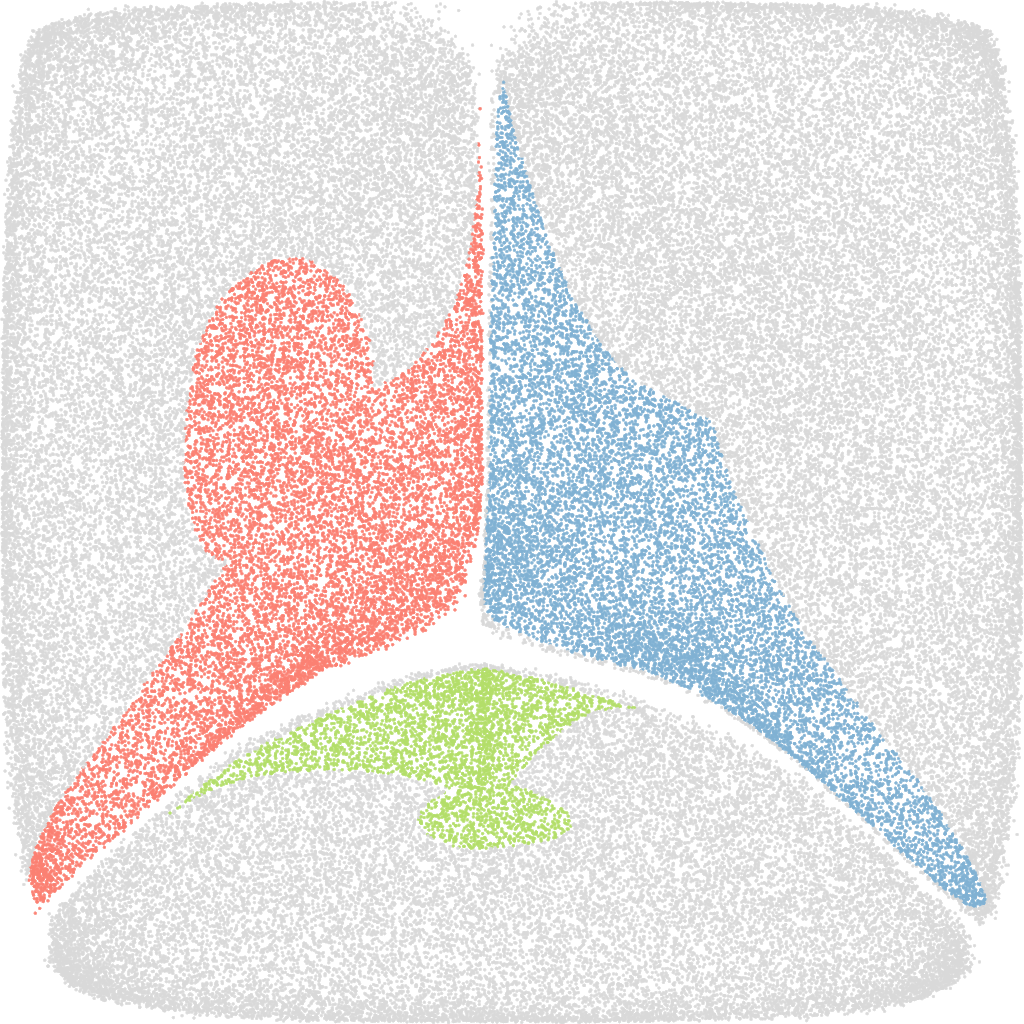} \label{fig::mix_16}
\caption{De-cluttering scatterplot with manual selections of different shapes highlighted by color. Selected regions change their geometry during the regularization iterations but preserve their sharp boundaries. No mixing of highlighted and other samples occurs. \editMinor{}{Left: Original scatterplot. Middle: After 2 iterations. Right: After 16 iterations.}}
\label{fig::mix}
\end{figure}

\renewcommand{\mysize}{0.4}
\begin{figure}[!hbt]
\centering
    \subfloat[\editMinor{original, 2.92ms}{}]{\includegraphics[width=\mysize\linewidth]{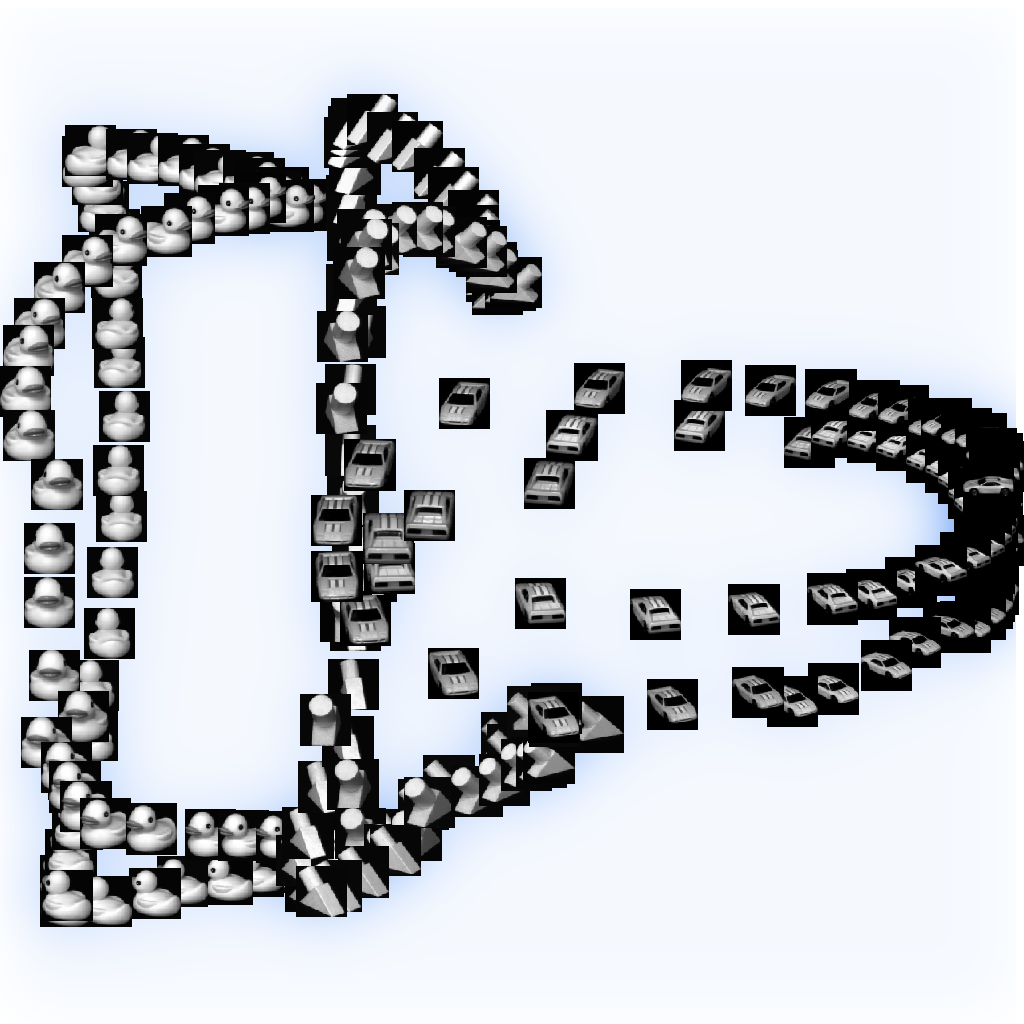} \label{fig::teaser_orig}}\quad
    \subfloat[\editMinor{2 iterations, 4.36ms}{}]{\includegraphics[width=\mysize\linewidth]{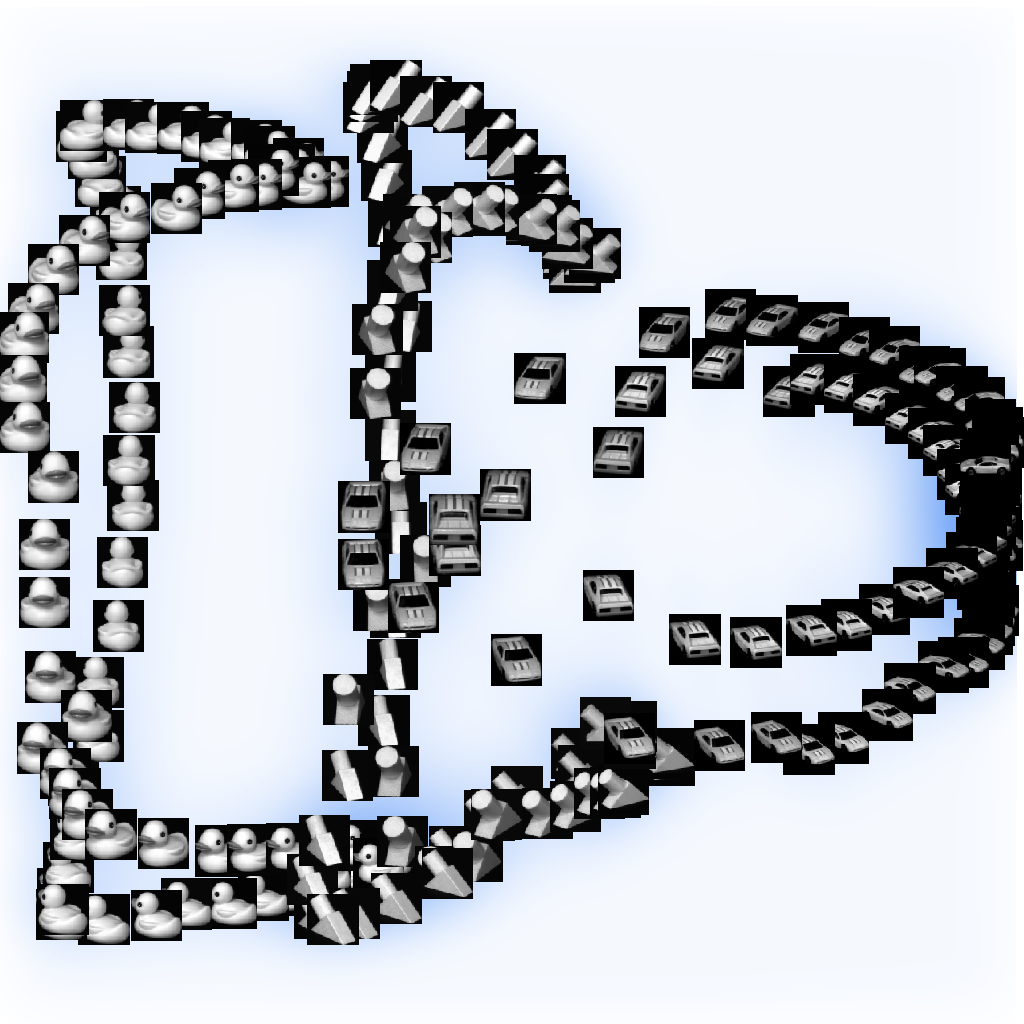} \label{fig::teaser_2}}\quad
    \subfloat[\editMinor{8 iterations, 13.50ms}{}]{\includegraphics[width=\mysize\linewidth]{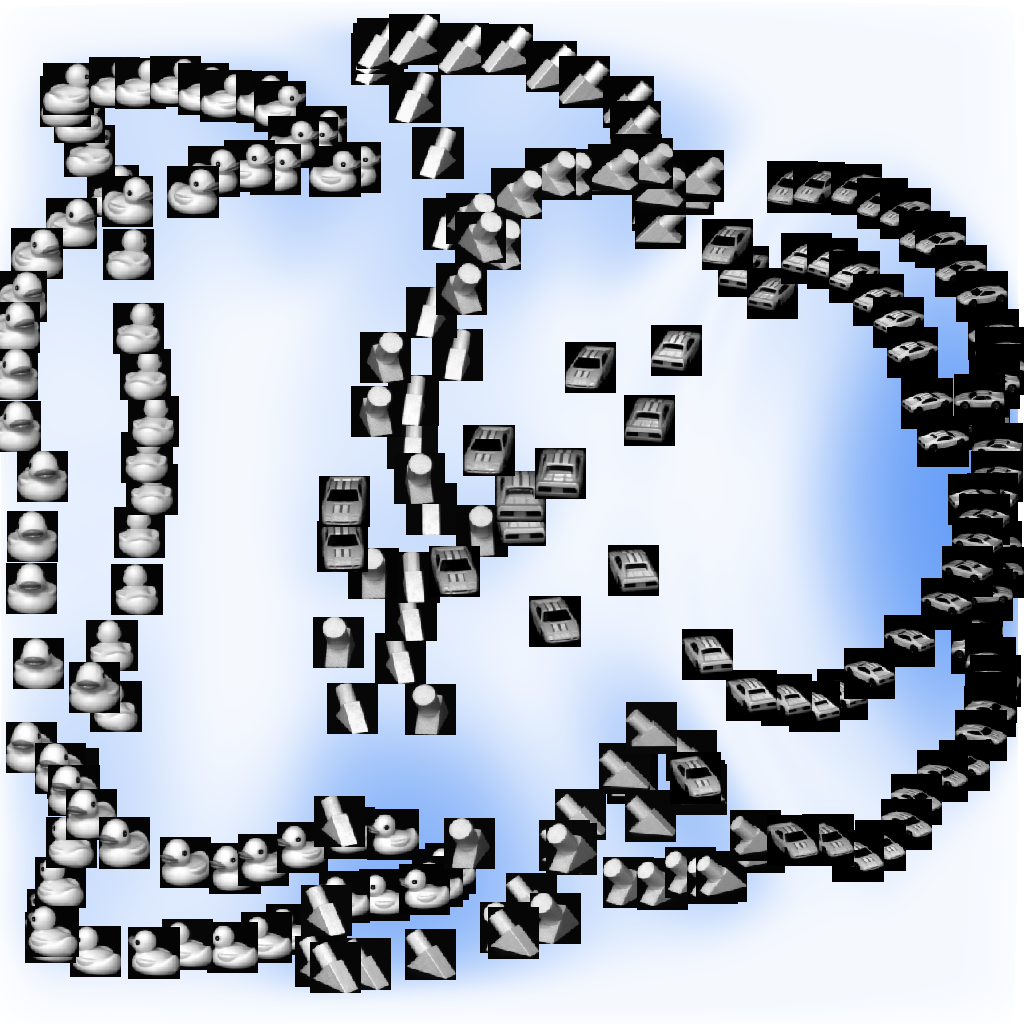} \label{fig::teaser_8}}\quad
    \subfloat[\editMinor{16 iterations, 26.37ms}{}]{\includegraphics[width=\mysize\linewidth]{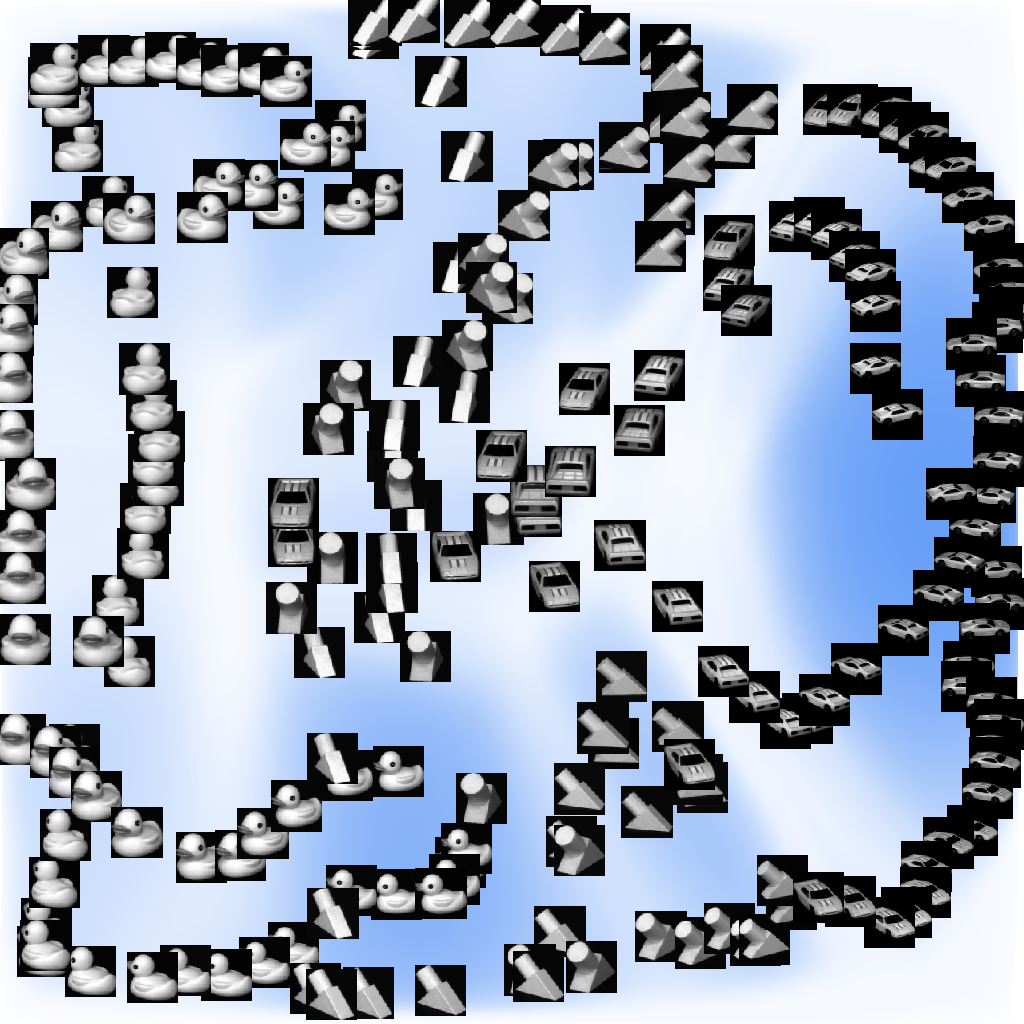} \label{fig::teaser_16}}
\centering
   \caption{De-cluttering scatterplot of COIL dataset~\cite{COIL96} with image icons. After a few iterations of our regularization, the user has a better overview of the variability of the data within clusters and can more easily access individual samples, as the screen space is used more effectively. The background texture encodes the original density of samples to reveal cluster structures after deformation. \editMinor{}{(a) Original, $2.92$~ms. (b) 2 iterations, $4.36$~ms. (c) 8 iterations, $13.50$~ms. (d) 16 iterations, $26.37$~ms}.}\label{fig:teaser}
\end{figure}

For some applications, it is desirable to replace point renderings in scatterplots with glyphs or icons, which require more space. Figure~\ref{fig:teaser} shows a scatterplot (with background coloring) of an image data collection with image icons. De-cluttering of the original layout results in a significant reduction of the icons' overlap using an efficient use of the available screen space. Therefore, the user can better inspect the variability of samples within clusters, access individual samples, estimate the density and numbers of data samples, and analyze class-cluster relations.

\subsection{Numerical Tests}
\label{sec:numerical_tests}
 
\smallskip
\noindent
\textbf{Performance and quality measures.} Table~\ref{tab:performance} shows that the \emph{computation time} scales linearly with the number of points and the number of iterations performed. Computation times are small enough for embedding our approach into interactive visual systems, even for large data sets.

\begin{table}[!hbt]
    \caption{Computation times on a GeForce RTX 2060 for a texture size of 1024$\times$1024 and a kernel size of $r=8$.}
    \label{tab:performance}
    \centering
    \begin{tabular}{|l|c|c|c|c|} \hline
        iter.
        & $500k$ & $1M$ & $2M$ & $4M$ \\ \hline
    
        1&
        1.16 ms&1.44 ms&1.98 ms&3.07 ms\\
        2&2.33 ms&2.95 ms&4.12 ms&6.40 ms\\
        3&3.56 ms&4.51 ms&6.33 ms&9.89 ms\\
        4&4.78 ms&6.06 ms&8.53 ms&13.33 ms\\
        5&6.06 ms&7.66 ms&10.81 ms&17.05 ms\\
        6&7.34 ms&9.30 ms&13.16 ms&20.81 ms\\
        7&8.63 ms&10.96 ms&15.56 ms&24.67 ms\\
        8&9.92 ms&12.63 ms&17.97 ms&28.55 ms\\ \hline
    \end{tabular}
\end{table}

\renewcommand{\mysize}{0.48}
\begin{figure}[!hbt]
  \centering
\includegraphics[width=\mysize\linewidth]{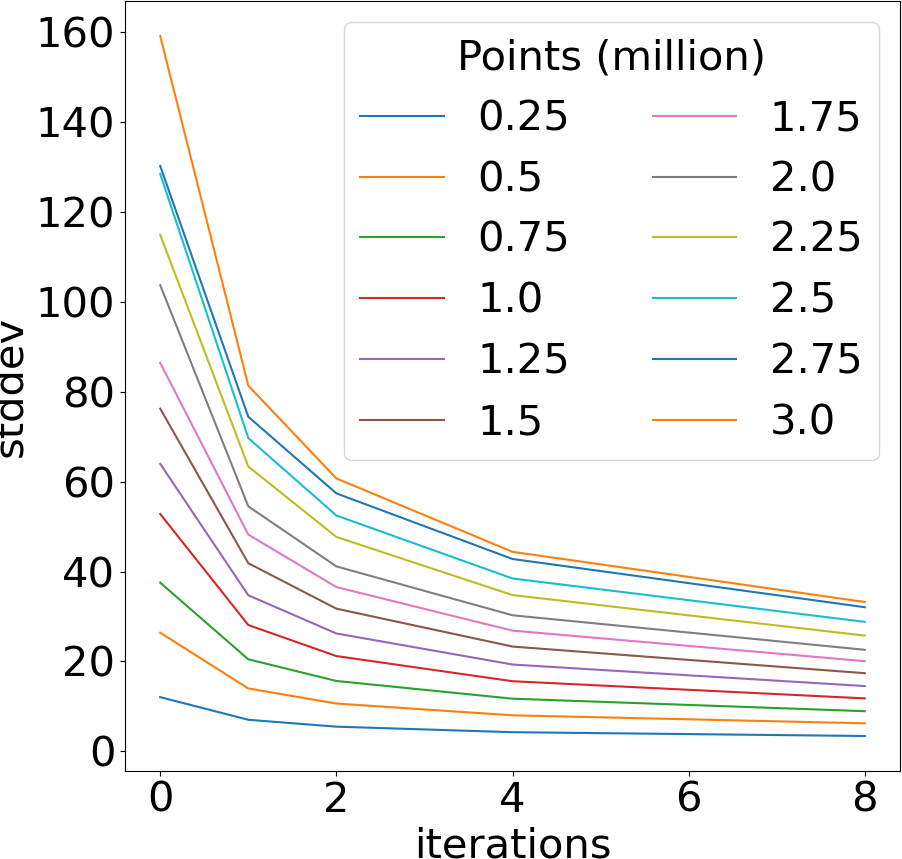}\ 
\includegraphics[width=\mysize\linewidth]{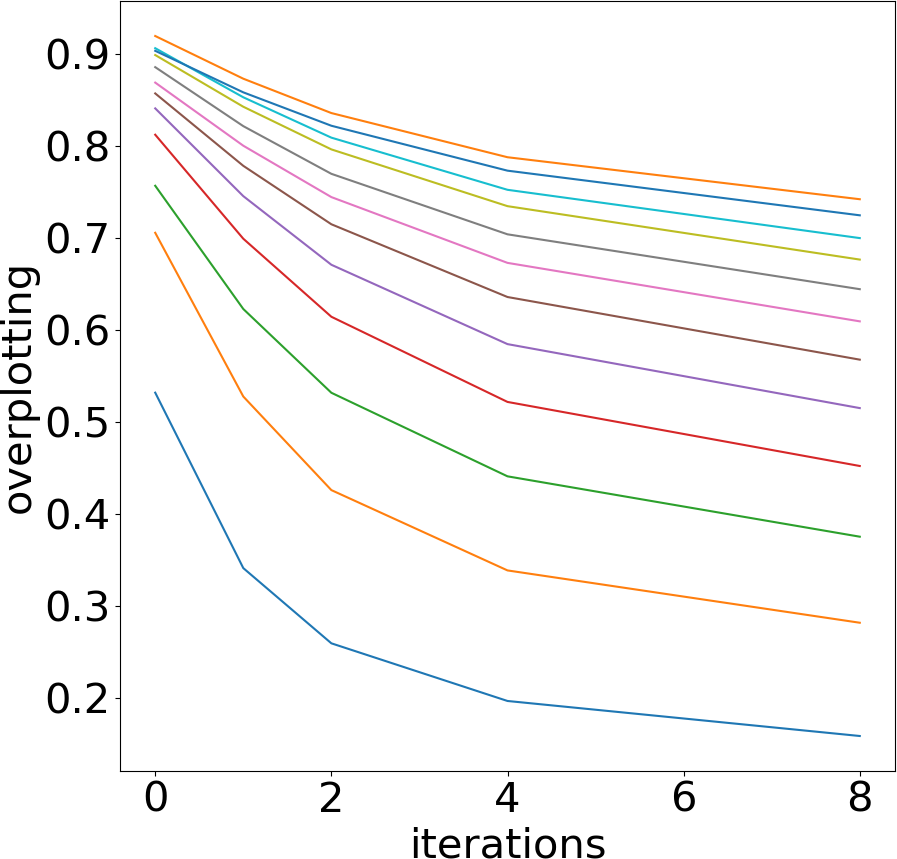}
\caption{Quality and performance measures of the proposed de-cluttering algorithm. The lowest curve (blue) corresponds to $250k$ samples. For each consecutive curve, the number of points is increased by $250k$ such that the upper-most curve (orange) corresponds to $3M$ points. Standard deviation from the mean number of samples and overplotting decrease monotonically, i.e., the samples' density becomes uniform. 
}
\label{fig::statistics}
\end{figure}

To judge the \emph{quality of our regularization}, we split the domain into bins of size $4\times 4$ pixels and compute the number of samples in each of these bins. Then, the standard deviation of the computed values from the mean number of samples per bin can serve as a regularization measure. The standard deviation vanishes for a perfectly regular sample distribution. Results presented in Figure~\ref{fig::statistics}, left show a monotonic decrease of the observed quantity over the iterations, i.e., the sample distribution becomes more and more regular. 

We furthermore compute the amount of overplotting as the difference between the total number of samples and the number of occupied pixels divided by the total number of samples.
Figure~\ref{fig::statistics}, right shows that overplotting monotonically decreases throughout the iterative de-cluttering. To compute binning and overplotting measures, we generated $500$ random datasets containing $1$ to $8$ variable-sized Gaussian clusters at random locations similar to~\cite{Mayorga13}.

\smallskip
\noindent
\textbf{Comparison with the state of the art.} \textit{Generalized Scatterplots} proposed by Keim et al.~\cite{Keim10} use linear distortions in horizontal and vertical directions similar to the \textit{HistoScale} method~\cite{Keim03} enhanced with a pixel-placement procedure. The linear distortion is the most similar technique to our method. Figure~\ref{fig::func_diag} shows an example of a data set that cannot be de-cluttered using the HistoScale approach. Thus, our regularization algorithm performs significantly better than the axis-aligned transformation of HistoScale. Figure~\ref{fig::reg} shows a visual comparison of our approach and the Generalized Scatterplots (for different parameter settings). Our technique uses the available screen space significantly better, which reduces occlusion. Moreover, in contrast to Generalized Scatterplots, our approach avoids overlap of the mapped data and does not require setting up related parameters.

\textit{Opacity adjustment} using $\alpha$-blending is arguably the most popular technique for mitigating visual clutter in scatterplots. Since this approach falls into the category of appearance change instead of spatial transformation, an objective comparison is less obvious. Therefore, we extensively compare our proposed regularization technique with opacity-adjusted scatterplots in our user study. Results are presented in Section~\ref{sec:user_study}.

We compared the results of the proposed de-cluttering algorithm with \edit{}{the state-of-the-art approach for } relaxed scatterplots developed by Raidou et al.~\cite{Raidou19}. We used the same datasets as the authors and refer the reader to the supplementary material for details on the produced visualizations. \edit{}{As datasets A through H were not available to us, we recreated them as closely as possible based on the information in their paper. } First, we observe that, if the domain resolution is fixed, the pixel-based mapping proposed by Raidou et al. cannot handle datasets with the number of samples exceeding the number of pixels\edit{, while our method can handle them.}{. Although the canvas size can be increased to satisfy the condition, a large canvas size may potentially exceed the maximal texture size on the GPU, which would hinder its use for fast computations.
In our proposed approach, such a limitation does not exist: The texture size affects only the quality of the resulting regularization and can always be chosen to match the GPU characteristics. Datasets of arbitrary sizes can be processed since the point-based data is converted into a density distribution rasterized on the given canvas. } \edit{Then }{Moreover}, while we are preserving neighborhoods, the method by Raidou et al. does not guarantee any preservation of the neighborhood relations (see Datasets G and H in supplementary material), since it aims at minimizing displacements of samples. \edit{Thus, their method tends to avoid the contraction of low-populated regions in contrast to our algorithm. Their pixel-based mapping spreads high-density clusters over neighboring unoccupied regions (see Dataset B). }{} Overall their deformation pattern (displacement vector field) is much more irregular than the one in our approach. For example, close samples can be shifted to random directions, which is especially noticeable in low-density regions. Instead, our proposed algorithm preserved samples' local relations even in extreme cases, see Figure~\ref{fig::D}.

Another state-of-the-art overlap-reducing algorithm, Hagrid, was developed by Cutura et al.~\cite{Cutura21, Cutura22}. The authors made the source code available for testing and evaluation. We compare our approach against Hagrid\edit{. Recently, Hilasaca et al.~\cite{Hilasaca23} presented a consolidated list }{using a number } of quality metrics \edit{for overlap removing techniques when applied to }{proposed by Hilasaca et al.~\cite{Hilasaca23} for } scatterplots with glyphs. Out of those metrics, \textit{glyphs' overlap} and \textit{layout spread} do not apply to scatterplots with no glyphs. \textit{Stress} and \textit{displacement} implicitly depend on the glyphs' size: When glyphs are small enough, short displacements are sufficient to remove overlaps. Therefore, it is not directly possible to apply these metrics to our proposed algorithm, which is designed to reproduce a globally uniform distribution of samples with no glyphs. \textit{Aspect ratio} is trivially conserved in our application case. However, \textit{trustworthiness}~\cite{Venna01} and \textit{orthogonal ordering}~\cite{Misue95} evaluate the preservation of local neighborhoods and relative positions of regularized samples, which we visually confirmed for our proposed algorithm, e.g., in Figure~\ref{fig::mix}. \editMinor{\edit{In addition }{Additionly}}{Additionally}, we now present results for the mentioned quality metrics when applied to \edit{$2.832$ }{$2,832$ } scatterplot layouts generated using numerical attributes of datasets from UCI repository~\cite{Dua19} (see supplementary material for technical details) in Figure~\ref{fig::stat}. 
Our proposed algorithm (denoted as InIm) demonstrates significantly better results in preserving local neighborhoods and relative positions of samples when compared to Hagrid. Hagrid tends to break these properties when resolving collisions along the space-filling curves.

\renewcommand{\mysize}{0.48}
\begin{figure}[!hbt]
  \centering
  \includegraphics[width=\mysize\linewidth]{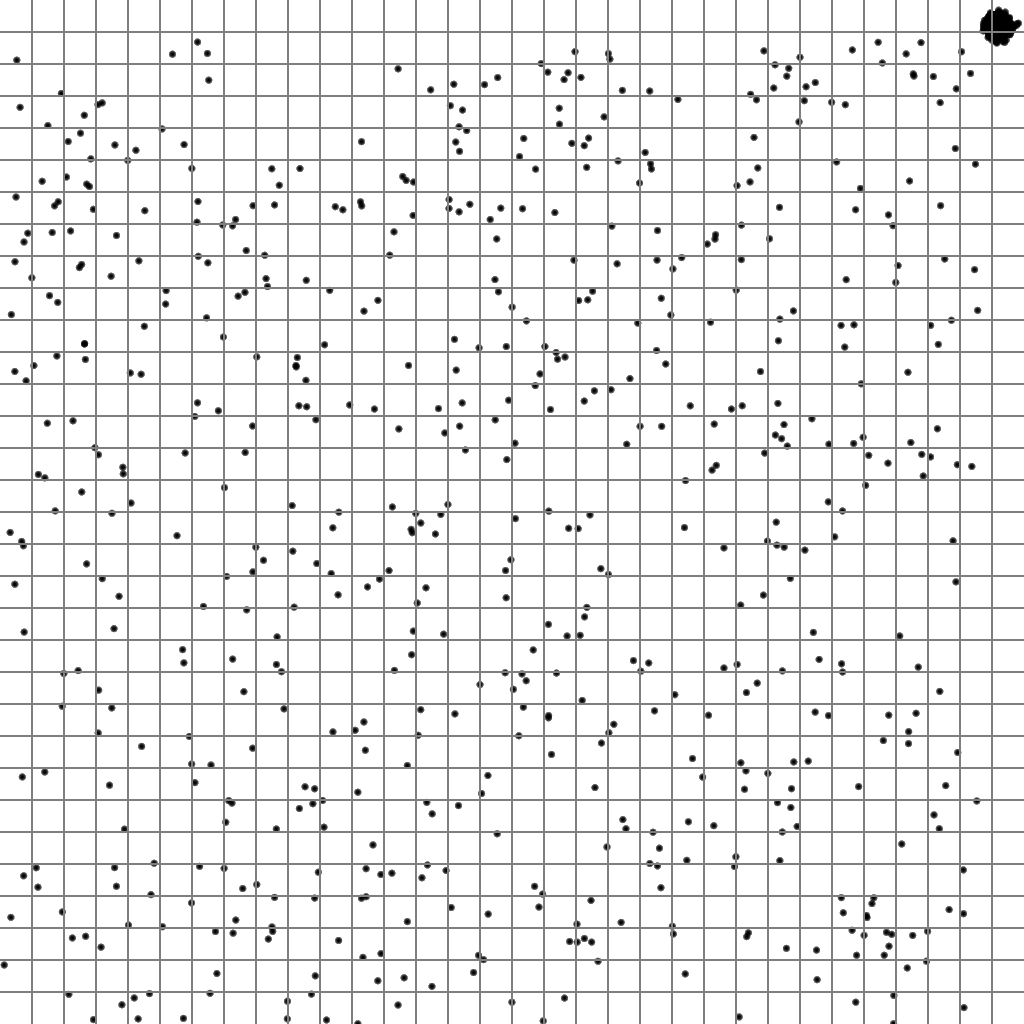} \label{fig::D0}\ 
  \includegraphics[width=\mysize\linewidth]{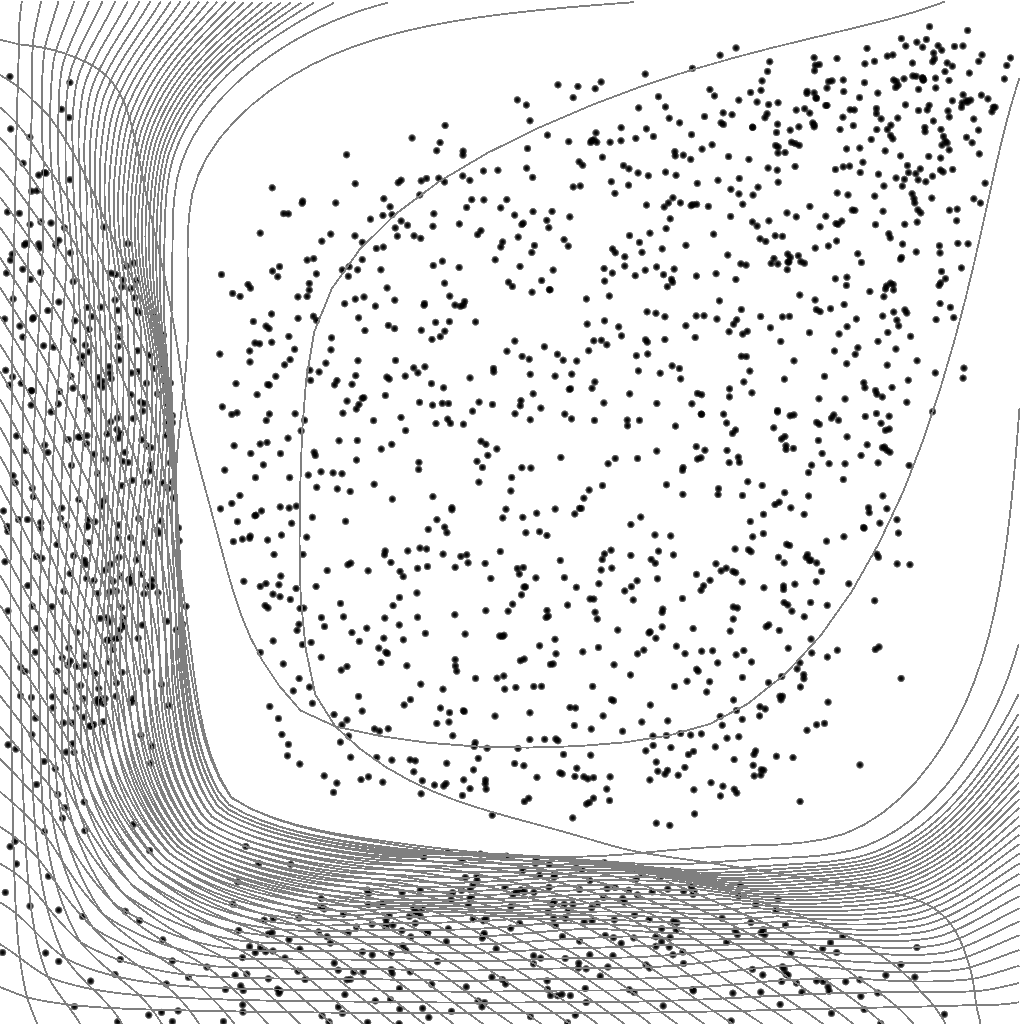} \label{fig::D1}
\caption{De-cluttering of \edit{}{the recreated }Dataset D from Raidou et al.~\cite{Raidou19} using our proposed algorithm. \editMinor{}{Left: Original scatterplot. Right: After 4 iterations.}}
\label{fig::D}
\end{figure}

\renewcommand{\mysize}{0.95}
\begin{figure}[!t]
  \centering
  \includegraphics[width=\mysize\linewidth]{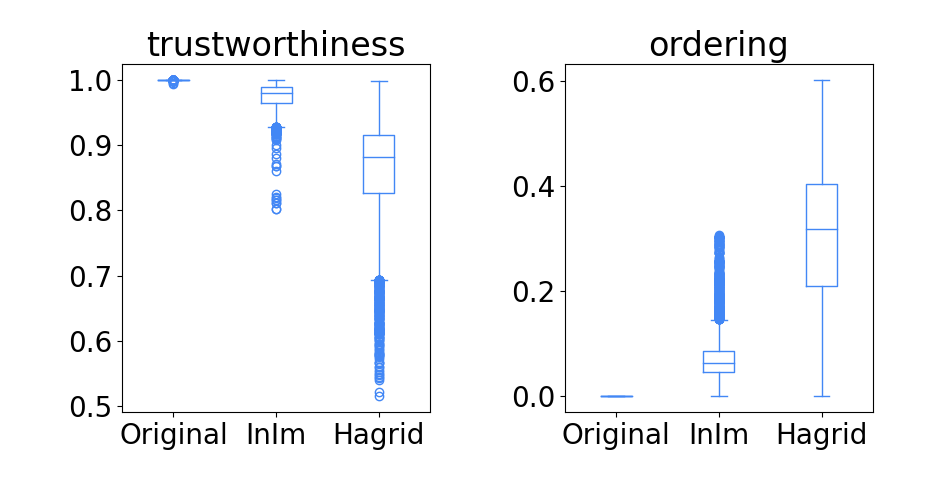}
\caption{Numerical comparison of our proposed method (InIm) with Hagrid using \edit{$2.832$ }{$2,832$ } scatterplots. Quality metrics \textit{trustworthiness} and \textit{ordering} characterizing the preservation of local data structure are evaluated. Our algorithm (referred to as InIm) results in significantly better values than Hagrid.
}
\label{fig::stat}
\end{figure}

\edit{}{The \textit{complexity} of the pixel-based relaxation algorithm by Raidou et al.~\cite{Raidou19} is $\mathcal{O}(n^3)$, which limits its application to large datasets. Therefore, the authors proposed to use an approximating median-split point-to-pixel mapping, which reduces the complexity to $\mathcal{O}(n\log{n})$ at the cost of worse preservation of the points' locality. The complexity of the collision handling part of Hagrid algorithm~\cite{Cutura22} is $\mathcal{O}(n^2)$ (with $\mathcal{O}(n\log n)$ operations in the best-case scenario). Our algorithm has a linear asymptotic complexity in the number of samples $n$ plus $\mathcal{O}(m)$ operations performed on the density texture with $m=2^k\times2^k$ pixels to compute InIms, leading to an overall complexity of $\mathcal{O}(n+m)$. Table~\ref{tab:performance} shows interactive performance rates even for datasets two orders of magnitude larger than the biggest datasets used in~\cite{Raidou19} and~\cite{Cutura22}.}

\smallskip
\noindent
\textbf{Smoothing parameter and background density value.} Smooth density fields ensure a smooth transformation of the scatterplot domain. The density smoothness is controlled by the kernel size $r$. Larger values of $r$ result in more regular deformations as shown in Figure~\ref{fig::regularization}. Our experiments showed that the results are robust against small changes of~$r$. A sufficiently large $r$ can, therefore, be set as default and does not require manual parameter tuning.

Our tests showed that vanishing density can lead to overlapping of mapped regions. Very large kernel sizes could alleviate the issue. However, we instead proposed to use the additive constant density introduced in~\ref{d0}, which eliminates vanishing density regions and therefore effectively prevents overlapping regions. This is a better solution than using very large kernel sizes as large values of $r$ lead to longer computation times as well as slightly slower convergence of the iterative regularization.

 Parameter $k$ defining the texture size should be large enough to achieve good sample separations in the regularized layout. However, unreasonably high values of $k$ result in increased consumption of memory and massively worsen the performance.

\renewcommand{\mysize}{0.46}
\renewcommand{\mysize}{0.4}
\begin{figure}[!tb]
  \centering
  \includegraphics[width=\mysize\linewidth]{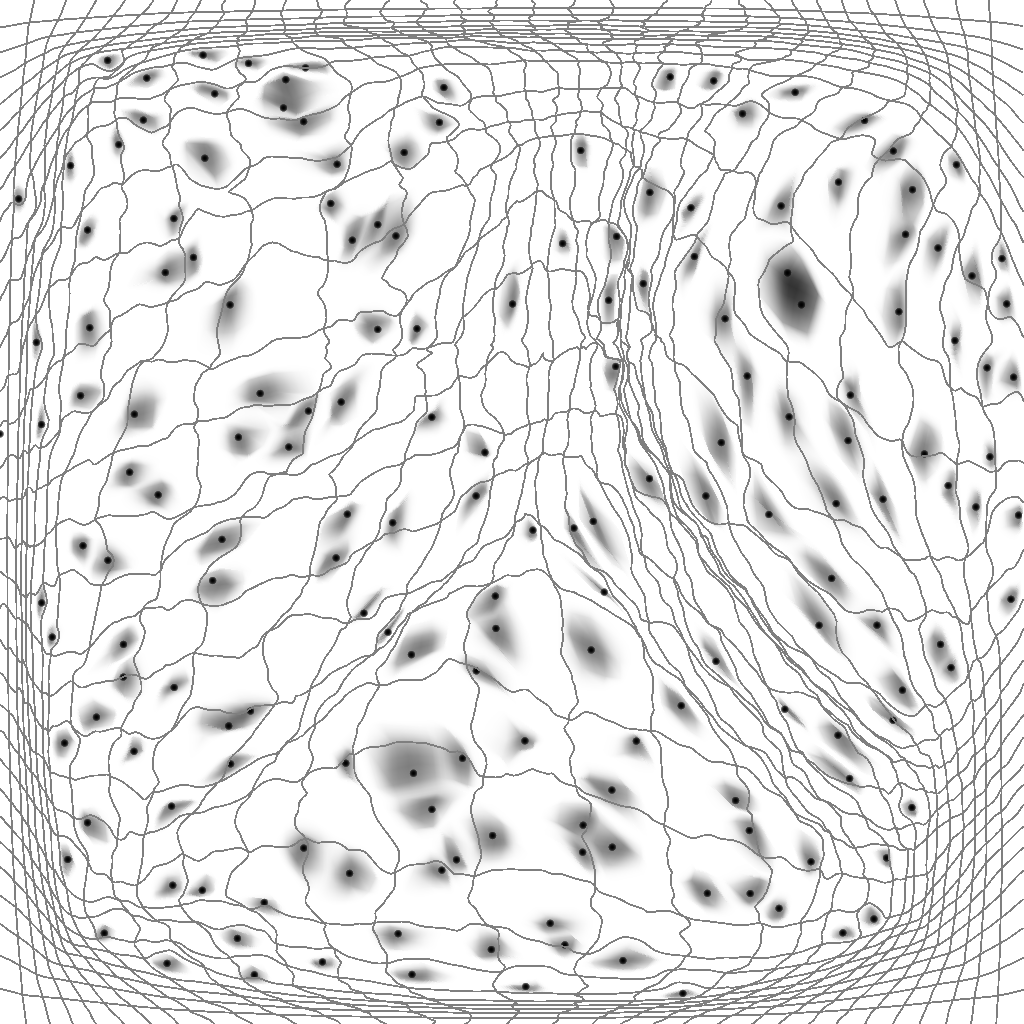} \label{fig::small_kernel}\quad
  \includegraphics[width=\mysize\linewidth]{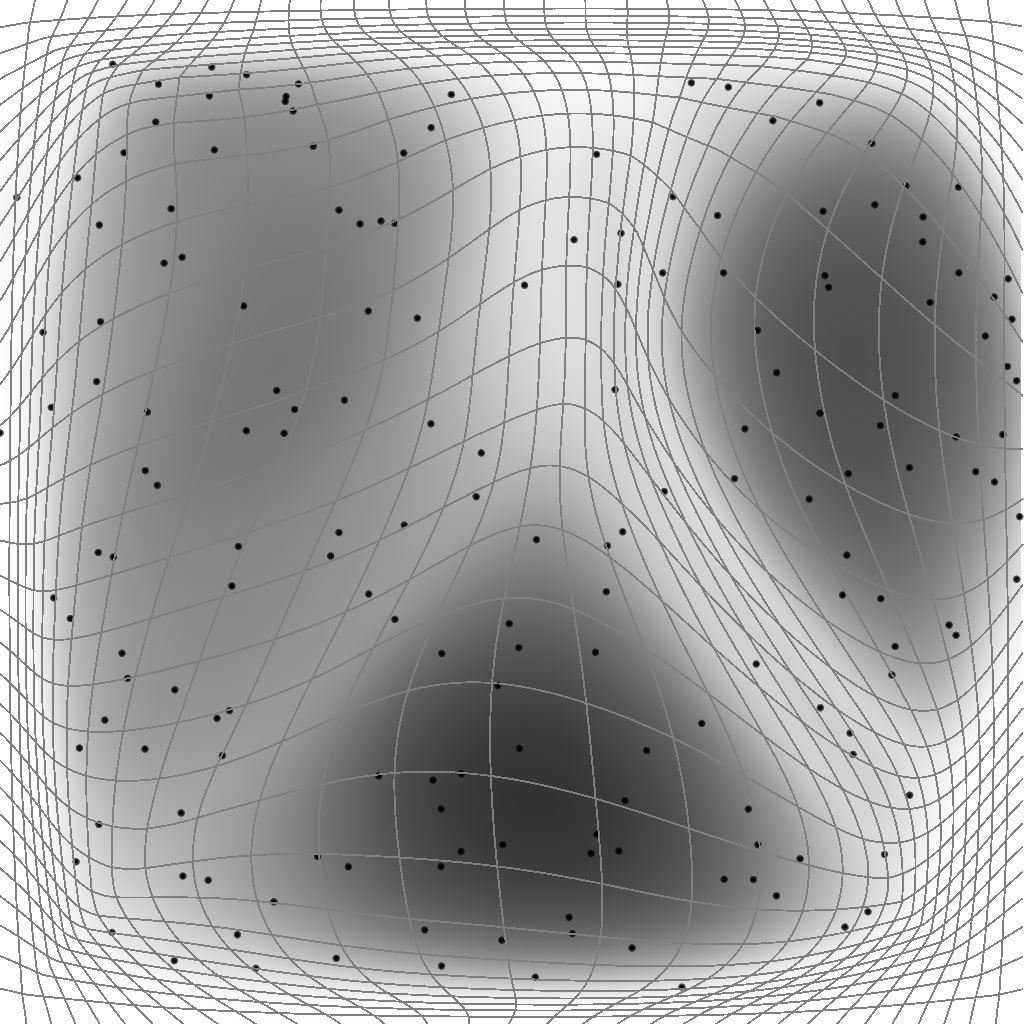} \label{fig::large_kernel}
  \\
\caption{Effect of varying size of the smoothing kernel on regularizing deformation for ``Wine'' dataset from UCI machine learning repository~\cite{Dua19}.
\editMinor{}{Left: }Very small kernel size results in an insufficiently smooth density distribution, which leads to a low quality of the mapping. Wriggled grid lines indicate the bad character of the mapping. \editMinor{}{Right: }Larger values of parameter $r$ improve the smoothness of the density and result in a well-behaved regularizing transformation.}
\label{fig::regularization}
\end{figure}

\subsection{User Study}\label{sec:user_study}

We conducted a quantitative user study to evaluate our approach against classical scatterplots with opacity adjustment, which is arguably the most commonly used method for clutter reduction in scatterplots. The goal of the study is to demonstrate that regularization may help to solve or mitigate issues related to occlusion (e.g., estimation of local number of samples, analysis of class-cluster interplay, accessibility of individual samples), while it is possible to overcome difficulties arising in the distorted layout (e.g., lack of clear cluster separation, equalized inter-sample distances) by suitable visual encodings of the distortion.

\noindent
\textbf{Setup.}
A total of 25 participants (aged 20 to 58 with an average of 28.56, 21 male, 3 female, 1 diverse) were recruited \edit{}{among friends and colleagues } to perform three different tasks on six datasets each. For each dataset, we measured the participants' accuracy, speed, and confidence. For the first two tasks, we compared two visualizations: the original scatterplot with density encoding and our regularized scatterplot. For the third task, we compared three visualizations: the original scatterplot, our regularized scatterplot with a grid, and our regularized scatterplot with a background texture. The participants were \edit{}{randomly } split into equally sized groups for each task. \edit{}{The study was conducted online and was fully anonymous. Before each task, the }\edit{The }{}visualizations and interaction mechanisms were explained in detail. \editMinor{}{All participants were, generally, familiar with scatterplots, but not all of them were visualization experts. Therefore, participants could familiarize themselves with the effects of our regularization approach on scatterplots and the different visual encodings by interactively changing the number of regularization iterations on an example scatterplot in a short training set-up. For the selection of points using a lasso selection tool, we had the participants perform a small test to make sure that interactions were correctly understood. }Each session took approximately 25 minutes.

\noindent
\textbf{Datasets and tasks.}
All $18$ datasets used for the user study are synthetic. Their visualizations are shown in the supplementary material. The participants were asked to complete the following three analysis tasks on the described datasets:
\begin{enumerate}
    \item[T1] \textit{Estimation of relative cluster sizes: } Given a scatterplot with two separate clusters, the participants were tasked to estimate what percentage of all points is in one of the clusters. The accuracy was measured using the absolute difference to the correct percentage. \edit{}{The clusters had the same shape to avoid unwanted visual effects that could affect the estimation of their sizes.}
    \item[T2] \textit{Sorting of clusters: } Given a scatterplot with multiple colored clusters, the participants were tasked to sort them by their number of points. The accuracy was measured by the number of correct pairwise relative positions in the ordering. \edit{}{All clusters represent two-dimensional Gaussian distributions with different amounts of variance. Their positions in the scatterplot were randomized.}
    \item[T3] \textit{Selection of clusters: } Given a scatterplot with multiple clusters, the participants were tasked to select the clusters using a provided lasso selection tool. The accuracy was measured by the percentage of correctly selected samples. \edit{}{This task included datasets with more complex cluster shapes. In particular, non-convex clusters such as arcs.}
\end{enumerate}

\noindent
\editMinor{}{These tasks cover the three task categories \textit{object-centric}, \textit{browsing}, and \textit{aggregate-level} for scatterplot designs proposed by Sarikaya and Gleicher~\cite{Sarikaya18}. More precisely, the tasks \textit{object comparison (4)}, \textit{search for known motif (6)}, \textit{characterize distribution (8)}, and \textit{numerosity comparison (11)} are included.}

\smallskip
\noindent
\textbf{Hypotheses.} We formulate the following hypotheses:
\begin{enumerate}
    \item[H1] Visual analysis of the data class-cluster composition is more precise when using our proposed approach compared to the classical scatterplots with density-based opacity adjustment \edit{}{(T2)}.
    \item[H2] Estimation of the cluster size is more \edit{reliable }{accurate } when using our proposed algorithm compared to the classical scatterplots with density-based opacity adjustment \edit{}{(T1)}.
    \item[H3] \textcolor{black}{Detection \edit{}{and selection} of clusters is equally \edit{possible }{accurate } in classical scatterplots with density-based opacity adjustment and when using our regularization enhanced by visual encodings of local deformations \edit{}{(T3)}.}
    \item[H4] \textcolor{black}{Visual encodings of local deformations \edit{work equally well}{lead to equal accuracy} with deformed grids and background coloring \edit{}{(T3)}.}
\end{enumerate}

\noindent
\textbf{Statistical analysis.}
We tested the null hypothesis that all approaches perform equally well in accuracy, speed, and confidence.
For the statistical analysis, we look at the p-value, calculated using a two-sample unpooled t-test, and the effect size, calculated using Cohen's $d$~\cite{Cohen}. We consider the p-value to be significant if it is smaller than the significance level of $0.05$.

\noindent
\textbf{Results and discussion.}
All statistical information about our analysis is provided in the supplementary material.
For the first task, the regularized visualization performed significantly better than the original scatterplot with opacity adjustment in both accuracy ($p=0.008$, $d=-0.569$) and confidence ($p\edit{=0.000 }{ < 10^{-3}}$, $d=1.154$). We therefore accept H2.
For the second task, the regularized visualization performed significantly better than the original scatterplot with opacity adjustment in both accuracy ($p\edit{=0.000 }{ < 10^{-3}}$, $d=0.713$) and confidence ($p\edit{=0.000 }{ < 10^{-3}}$, $d=0.752$). We therefore accept H1.
For the third task, the regularized visualization with background texture performed significantly worse than the original scatterplot in accuracy ($p=0.048$, $d=-0.400$) and the regularized visualization with grid performed significantly worse than the regularized visualization with background texture in accuracy ($p\edit{=0.000 }{ < 10^{-3}}$, $d=-1.142$), speed ($p=0.027$, $d=0.452$), and confidence ($p\edit{=0.000 }{ < 10^{-3}}$, $d=-1.053$). We therefore reject H3. We also reject H4, as the background texture visualization performed significantly better than the deformed grid. \editMinor{}{A limitation of our study is the exclusive use of Gaussian clusters for Task T2. More complex cluster shapes would have made the task more difficult for participants and possibly caused unintended side effects. Moreover, since our approach does not preserve the shape of clusters, we expect the same results for non-Gaussian clusters.}

\section{Conclusion}\label{sec:conclusion}

We proposed an algorithm for a data-driven deformation of the visual domain, which can be interactively toggled by the user to de-clutter the scatterplot layout. The algorithm is deterministic, computes progressive regularizations of the initial layout, and has theoretical and practical complexity of \edit{$\mathcal{O}(n)$ }{$\mathcal{O}(n+m)$ } as it does not \edit{reuire }{require } any collision detection of samples. Thus\edit{, }{ } the user can change the desired degree of regularization applied to large datasets in run time. An efficient GPU implementation is provided as an open-source code. The resulting density-equalized layouts preserve original local relations of data items and allow for better accessibility of samples, accurate selections, and easier analysis of class-cluster characteristics. Although some data analysis tasks are better performed in the original scatterplot, principal information about the data structure can be still visually presented in the de-cluttered layout, which can be beneficial for the user. We investigated several approaches for visual encoding of the deformation map including grid, density texture, and contour lines. \edit{}{While grid lines better characterize the resulting deformation, density texture helps to identify data clusters and contour lines reveal subcluster structures, which may be occluded in the original plot.}

The choice of parameters (additive constant density, kernel size) is discussed and visually illustrated. Computational efficiency, preservation properties and application scenarios of the proposed technique were visually and numerically explored using about \edit{$3.400$ }{$3,400$ } different scatterplot layouts. We compared our method to four existing approaches \cite{Keim03, Keim10, Raidou19, Cutura21} by constructing representative examples and evaluating important quality metrics. Detailed information about our tests is presented in the supplementary material and the accompanying video.

\edit{}{Most of the existing regularization approaches compute the displacement of each sample based on local data distribution. Frequently, it leads to collisions when several samples are moved to the same location. Such collisions should be then detected and resolved, making the overall algorithm complex and slow. In contrast, the proposed technique determines individual displacements of samples based on information about data distribution on the global scope. This information is encoded in a set of InIms, which can be efficiently computed. So, we ensure a monotonous decrease of samples' density in the whole spatial domain and avoid collisions, therefore achieving fast convergence to a nearly uniform sample distribution.}

Density-equalizing mappings find their application in various visualization tasks.
Future directions of research may include the application of the proposed technique to local lenses in scatterplots, general optimization of items' placement in layouts, and construction of contiguous cartograms.

\section*{Acknowledgments}
\edit{}{This work was funded by the Deutsche Forschungsgemeinschaft (DFG) – \mbox{MO 3050/2-3} and \mbox{CRC 1450 – 431460824}.}

\bibliographystyle{IEEEtran}
\bibliography{IEEEabrv, references.bib}

\begin{IEEEbiographynophoto}
{Hennes Rave}\editMinor{}{is a doctoral researcher in the Visualization and Graphics (VISIX) group at the University of M\"unster, Germany, where he received his Master's degree in Computer Science in 2021. His research interests include interactive visualization, spectral image visualization, and computer graphics.}
\end{IEEEbiographynophoto}

\vspace{-2.95em}

\begin{IEEEbiographynophoto}
{Vladimir Molchanov}\editMinor{}{is a post-doctoral researcher with University of M\"unster, Germany. He received his Bachelor's and Master's degrees in Applied Mathematics from Novosibirsk State University, Russia, and his Ph.D. degree in Mathematics from Jacobs University, Bremen, Germany. His research interests include projection methods, interactive exploratory systems, visualization of multidimensional data, and biomedical data analysis.}
\end{IEEEbiographynophoto}

\vspace{-2.95em}

\begin{IEEEbiographynophoto}
{Lars Linsen}\editMinor{}{is a Full Professor of Computer Science at the University of M\"unster, Germany. He received his academic degrees from the University of Karlsruhe, Germany, including a Ph.D. in Computer Science. Subsequent affiliations were the University of California, Davis, U.S.A., as a post-doctoral researcher and lecturer, the University of Greifswald, Germany, as an assistant professor, and Jacobs University, Bremen, Germany, as an associate and full professor. His research interests are in interactive visual data analysis.}
\end{IEEEbiographynophoto}

\vfill

\end{document}